\documentstyle[referee,epsfig,url]{mn}
\input epsf
\newcommand{\kms}{\>{\rm km}\,{\rm s}^{-1}}

\newcommand{\kpc}{\>{\rm kpc}}
\newcommand{\Msun}{\>{\rm M_{\odot}}}
\newcommand{\Lsun}{\>{\rm L_{\odot}}}

\newcommand{\photo}{\>{\rm\bf Photo}}
\newcommand{\target}{\>{\rm\bf target}}
\newcommand{\SB}{\>{\rm magarcsec^{-2}}}
\newcommand{\acrsec}{\char'175}
\newcommand{\Sersic}{S\'ersic }
\def\Log{\,{\rm Log}\,}
\def\ln{\,{\rm ln}\,}
\def\exp{\,{\rm exp}\,}
\newcommand{\reference}{\bibitem}
\newcommand{\beq}{\begin{equation}}
\newcommand{\eeq}{\end{equation}}

\def\kms{\,{\rm {km\, s^{-1}}}}

\begin{document}

\title [The size distribution of galaxies]
{The size distribution of galaxies in the Sloan Digital Sky Survey}
\author[Shen et al]
{Shiyin Shen$^{1,2}$,  H.J. Mo$^{1}$, Simon D.M. White$^{1}$, Michael R. Blanton$^{3}$,  \and
Guinevere Kauffmann$^{1}$,  Wolfgang Voges$^{2}$, J. Brinkmann$^4$, Istvan Csabai$^4$
\thanks {E-mail: shen@mpa-garching.mpg.de}
 \\
$^1$Max-Planck-Institut f\"ur Astrophysik, Karl Schwarzschilddirectory 
         Str. 1, Postfach 1317, 85741 Garching, Germany\\
$^2$ Max-Planck-Institut f\"ur extraterrestrische Physik, Postfach 1312,
85741 Garching, Germany \\
$^3$ Center for Cosmology and Particle Physics, Department of Physics,  New York University, 4 Washington Place, \\ New York,
NY 10003, USA\\
$^4$ Apache Point Observatory, P.O. Box 59, Sunspot, NM 88349, USA}

\maketitle


\begin{abstract}
We use a complete sample of about 140,000 galaxies from the
Sloan Digital Sky Survey (SDSS) to study the size distribution of
galaxies and its dependence on their luminosity, stellar mass, and
morphological type. The large SDSS database provides statistics of
unprecedented accuracy. For each type of
galaxy, the size distribution at given luminosity (or stellar
mass) is well described by a log-normal function, characterized by
its median $\bar{R}$ and dispersion $\sigma_{\ln R}$.
For late-type
galaxies, there is a characteristic luminosity at $M_{r,0}\sim
-20.5$ (assuming $h=0.7$) corresponding to a stellar mass $M_0\sim
10^{10.6}\Msun$. Galaxies more massive than $M_0$ have
$\bar{R}\propto M^{0.4}$ and $\sigma_{\ln R}\sim 0.3$, while less
massive galaxies have $\bar{R}\propto M^{0.15}$ and $\sigma_{\ln
R}\sim 0.5$. For early-type galaxies, the $\bar{R}$ - $M$ relation
is significantly steeper,  $\bar{R}\propto M^{0.55}$, but the
$\sigma_{\ln R}$ - $M$ relation is similar to that of bright
late-type galaxies. Faint red galaxies have sizes quite independent
of their luminosities.
We use simple theoretical models to interpret these
results. The observed $\bar{R}$ - $M$ relation for
late-type galaxies can be explained if the fraction of baryons that
form stars is as predicted by the
standard feedback model. Fitting the
observed $\sigma_{\ln R}$ - $M$ relation requires in addition that
the bulge/disk mass ratio be larger in haloes of lower angular
momentum and that the bulge material transfer part of its angular
momentum to the disk. This can be achieved
if bulge formation occurs so as to maintain a marginally stable disk.
For early-type galaxies the observed $\sigma_{\ln R}$ -
$M$ relation is inconsistent with formation through single
major mergers of present-day disks. It is
consistent with formation through repeated mergers, if the progenitors
have properties similar to those of faint ellipticals or Lyman break
galaxies and merge from relatively strongly bound orbits.
\end{abstract}

\section{Introduction}

Luminosity, size, circular velocity (or velocity
dispersion), and morphological type are the  most basic properties
of a galaxy. Observed galaxies cover large ranges in these
properties, with luminosities between $\sim 10^8 \Lsun$ and $\sim
10^{12} \Lsun $, effective radii between $\sim 0.1 \kpc$ and $\sim
10 \kpc$, circular velocity (or velocity dispersion) between $\sim
30\kms$ and $\sim 300\kms$, morphologies changing from pure disk
systems to pure ellipsoidal systems. Clearly, the study of the
distribution of galaxies with respect to these properties and the
correlation among them are crucial to our understanding of the
formation and evolution of the galaxy population.

There has been much recent progress in this area. For example, the
luminosity function of galaxies has been measured from various
redshift surveys of galaxies and is found to be well described by
the Schechter function (Schmidt 1968; Loveday et al. 1992; Lin et
al. 1996; Folkes et al. 1999; Madgwick et al. 2002);
the morphological type of galaxies
is found to be correlated with their local environment (as
reflected in the the morphology-density relation, Dressler 1980;
Dressler et al. 1997) and other properties (e.g. Roberts \& Haynes
1994); and galaxy sizes are correlated with luminosity and
morphological type (Kormendy 1977), and have a distribution which
may be described by a log-normal function
(Choloniewski 1985; Syer, Mao \& Mo 1999;
de Jong \& Lacey 2000).

Clearly, in order to examine these properties in detail, one needs
large samples of galaxies with redshift measurements and accurate
photometry. The Sloan Digital Sky Survey (SDSS, York et al. 2000),
with its high-quality spectra and good photometry in five bands
for $\sim 10^6$ galaxies, is providing an unprecedented database
for such studies. The survey is ongoing, but the existing data are
providing many interesting results about the distribution of
galaxies with respect to their intrinsic properties. The
luminosity function has been derived by Blanton et al. (2001,
2002c) and the dependence of luminosity function on galaxy type
has been analyzed by Nakamura et al. (2003). Shimasaku et al.
(2001), Strateva et al. (2002) and Nakamura et al. (2003) have
examined the correlation between galaxy morphological type and
other photometric properties, such as color and image
concentration. The fundamental-plane and some other
scaling relations of early-type galaxies have been 
investigated by Bernardi et al. (2003a, 2003b, 2003c). Based on a
similar data set, Sheth et al. (2003) have studied the
distribution of galaxies with respect to central velocity
dispersion of galaxies. By modelling galaxy spectra in detail,
Kauffmann et al. (2002a) have measured stellar masses for a sample
of more than $10^5$ galaxies, and have analyzed the correlation
between stellar mass, stellar age and structural properties
determined from the photometry (Kauffmann et al. 2002b). Blanton
et al. (2002b) have examined how various photometric properties of
galaxies correlate with each other and with environment density.

In this paper, we study in detail the size distribution of
galaxies and its dependence on galaxy luminosity, stellar mass and
morphological type. Our purpose is twofold: (1) to quantify these
properties so that they can be used to constrain
theoretical models; (2) to use simple models based on current
theory of galaxy formation to interpret the observations.
Some parts of our analysis overlap that of Blanton et
al. (2002b), Kauffmann et al. (2002b) and Bernardi et al. (2003b),
but in addition we
address other issues. We pay considerable attention to effects caused
by the sample surface brightness limit and by seeing, we provide
detailed fits to the data to quantify the dependence of the size
distribution on other galaxy properties, and we discuss how these
results can be modelled within the current
theory of galaxy formation.

Our paper is organized as follows. In Section 2, we describe
the data to be used. In Section 3, we derive
the size distribution of galaxies as a function of
luminosity, stellar mass, and type. In Section 4, we build
simple theoretical models to understand the observational
results we obtain. Finally, in Section 4, we summarize
our main results and discuss them further.

\section{The data}

In this section, we describe briefly the SDSS data used in this paper.
These data are of two types: the basic SDSS photometric and
spectroscopic data, and some quantities derived by our SDSS
collaborators from the basic SDSS database.

\subsection{The basic SDSS data}

The SDSS observes galaxies in five  photometric bands
($u,g,r,i,z$) centred at (3540, 4770, 6230, 7630, 9130$\rm\AA$)
down to 22.0, 22.2, 22.2, 21.3, 20.5 mag, respectively.
The imaging camera is described by Gunn et al. (1998);
the filter system is roughly as described in Fukugita et al. (1996);
the photometric calibration of the SDSS imaging data is described
in Hogg et al. (2001) and Smith et al. (2002); and the accuracy
of astrometric calibration is described in  Pier et al. (2003).
The basic SDSS photometric data base is then obtained
from an automatic software pipeline called $\photo$
(see Lupton et al. 2001, 2002), whereas the basic
spectroscopic parameters of each galaxy, such as
redshift, spectral type,  etc, are obtained by the spectroscopic
pipelines $\rm\bf idlspec2d$ (written by D. Schlegel \& S. Burles)
and $\rm\bf spectro1d$ (written by M. SubbaRao, M. Bernardi
and J. Frieman). Many of the
survey properties are described in detail in the Early Data
Release paper (Stoughton et al. 2002, hereafter EDR).

$\photo$ uses a modified form of the Petrosian (1976) system
for galaxy photometry, which is designed to measure a constant
fraction of the total light independent of the surface-brightness
limit. The Petrosian radius $r_P$ is defined to be the radius
where the local surface-brightness averaged in an annulus equals
20 percent of the mean surface-brightness interior to this annulus, i.e.
\begin{equation}\label {r_P}
 \frac{\int_{0.8r_P}^{1.25r_P}dr2\pi
rI(r)/[\pi(1.25^2-0.8^2)r^2]}{\int_{0}^{r_P}dr2\pi rI(r)/[\pi
r^2]}=0.2,
\end{equation}
where $I(r)$ is the azimuthally averaged surface-brightness
profile. The Petrosian flux $F_p$ is then defined as the total
flux within a radius of $2r_P$,
\begin{equation}\label{F_p}
F_p=\int_{0}^{2r_P}2\pi rdrI(r).
\end{equation}
With this definition, the Petrosian flux (magnitude) is about 98
percent of the total flux for an exponential profile  and about 80
percent for a de Vaucouleurs profile.  The other two Petrosian
radii listed in the $\photo$ output, $R_{50}$ and $R_{90}$, are
the radii enclosing 50 percent and 90 percent of the Petrosian
flux, respectively. The concentration index $c$, defined as $c
\equiv R_{90}/R_{50}$, is found to be correlated with galaxy
morphological type (Shimasaku et al. 2001; Strateva et al. 2002;
Nakamura et al. 2003). An elliptical galaxy with a de Vaucouleurs
profile has $c\sim 3.3$ while an exponential disk has $c\sim 2.3$.
Note that these $\photo$ quantities are not corrected for the
effects of the  point spread function (hereafter PSF) and,
therefore, for small galaxies under bad seeing conditions, the
Petrosian flux is close to the fraction measured within a typical
PSF, about $95\%$ for the SDSS (Blanton et al. 2001). In such
cases, the sizes of compact galaxies are over-estimated, while their
concentration indices are under-estimated by the uncorrected
$\photo$ quantities (Blanton et al. 2002b).

The SDSS spectroscopic survey aims to obtain a galaxy sample
complete to $r\sim 17.77$ in the $r$-band (Petrosian) magnitude
and to an average $r$-band surface-brightness (within $R_{50}$)
$\mu_{50} \sim 24.5 \SB$.  This sample is denoted the Main Galaxy Sample,
to distinguish it from another color-selected galaxy sample, the
Luminous Red Galaxies (LRGs), which extends to $r\sim 19.5$
(Eisenstein et al. 2001). These target selections are carried out
by the software pipeline $\target$; details about the target
selection criteria for the Main Galaxy Sample are described in
Strauss et al. (2002). The tiling algorithm of the fibers
 to these spectroscopic targets is described in Blanton et al. (2003).

The spectroscopic pipelines, $\rm\bf{idlspec2d}$ and
$\rm\bf{spectro1d}$, are designed to produce fully calibrated
one-dimensional spectra, to measure a variety of spectral features,
to classify objects by their spectral types, and to
determine redshifts.  The SDSS spectroscopic pipelines have an
overall performances such that the correct classifications and redshifts
are found for $99.7\%$
of galaxies in the main  sample (Strauss et al. 2002). The errors in
the measured redshift are typically less than $\sim 10^{-4}$.

\subsection{Derived quantities for SDSS galaxies}

In the SDSS $\photo$ output, the observed surface-brightness
profiles of galaxies are given in $\rm\bf {profMean}$ where
angle averaged surface-brightness in a series of annuli are listed (see
EDR). Blanton et al. (2002b) fitted the angle averaged  profiles
with the \Sersic (1968) model,
\begin{equation}\label{Sersic}
 I(r)=I_0\exp [-(r/r_0)^{1/n}],
\end{equation}
convolved with the PSF, to obtain the central surface-brightness
$I_0$, the scale radius $r_0$, and the profile index $n$ for each
galaxy. As found by many authors (e.g. Trujillo, Graham \& Caon
2001 and references therein), the profile index $n$ is correlated
with the morphological type, with late-type spiral galaxies (whose
surface-brightness profiles can be approximated by an exponential
function) having $n\sim 1$, and early-type elliptical galaxies
(whose surface-brightness profiles can be approximated by the
$r^{1/4}$ function) having $n\sim 4$. From the fitting results,
one can obtain the total \Sersic magnitude (flux), the \Sersic
half-light radius, $R_{50,S}$, and other photometric quantities.
In our following analysis, we will use these quantities and
compare the results so obtained with those based on the original
$\photo$ quantities. For clarity, we denote all the \Sersic
quantities by a subscript `$S$'. We present 
results for both Petrosian and \Sersic quantities, because
while the \Sersic quantities are corrected for seeing effect,
the Petrosian quantities are the standard photometric 
quantities adopted by the SDSS community.  

Recently, Kauffmann et al. (2002a) developed a method to estimate
the stellar mass of a galaxy based on its spectral features, and
obtained the stellar masses for a sample of 122,808 SDSS galaxies.
The $95\%$ confidence range for the mass estimate of a typical
galaxy is $\pm 40\%$. Below we use these
results to quantify size distributions as a function of
stellar mass as well as a function of luminosity.

\subsection{Our sample}

\begin{figure}
\epsfysize=14.0cm \centerline{\epsfbox{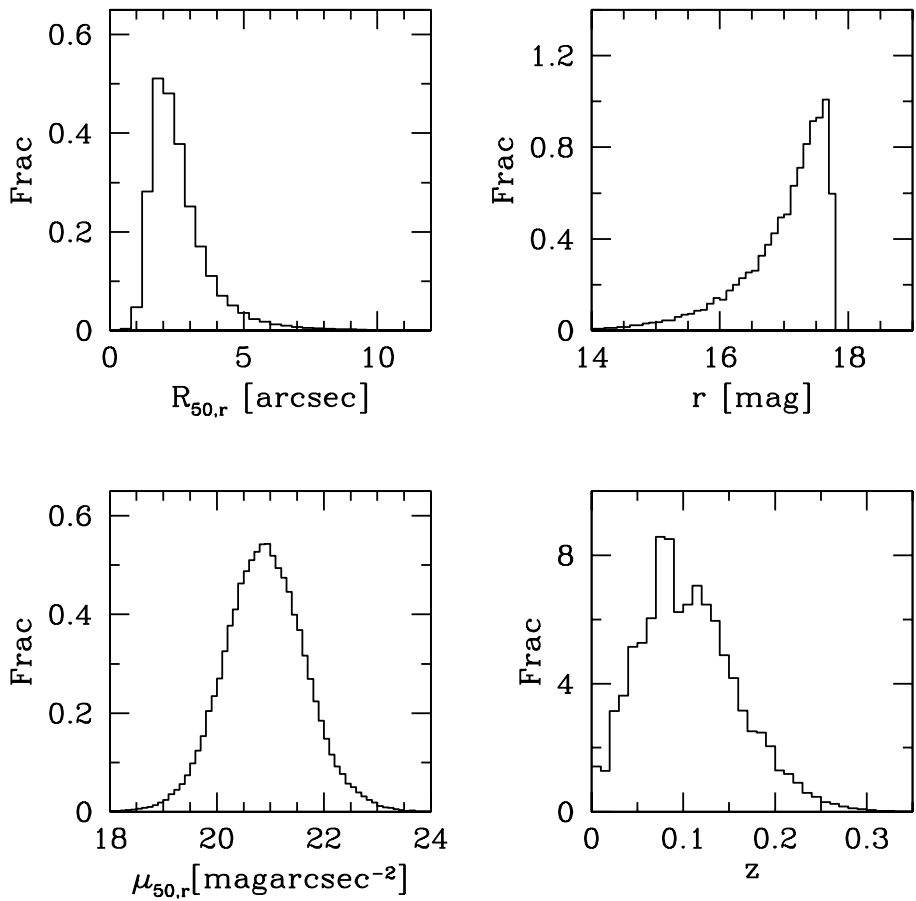}}
\caption{The distribution of galaxies with respect to
some basic SDSS photometric quantities and the redshift.
All histograms are normalized to 1.}
\label{histogram}
\end{figure}

The main sample we use in this paper is a sub-sample of the
spectroscopic targets  observed before April 2002, which is
known as the Large-Scale Structure (LSS) {\bf{sample10}} within
the SDSS collaboration (Blanton et al. 2002c).
We select from it 168,958 Main Galaxy
Targets (with SDSS flag TARGET$\_$GALAXY or
TARGET$\_$GALAXY$\_$BIG) with high-confidence redshifts
($\rm\bf{zWarning}=0$).

Figure 1 shows histograms of the basic quantities of our
selected sample. The top-right panel shows the galaxy distribution
in  $r$-band apparent magnitude $r$ after correction
for foreground Galactic extinction using the reddening map of
Schlegel, Finkbeiner and Davis (1998). The abrupt cut at
$\sim17.77$ mag is caused by the target selection criteria.  The
top-left panel shows the distribution of the $r$-band Petrosian
half-light radius $R_{50,r}$. The distribution of $\mu_{50,r}$, the $r$-band
average surface-brightness within $R_{50,r}$, is shown in the
bottom-left panel. Since the $r$-band is the reference band of the
SDSS for model fitting and target selection, our discussion of
sample incompleteness will be based on the photometric properties
in this band. The redshifts $z$ of the sample galaxies are obtained
from the spectroscopic data, and the distribution of galaxies with
respect to  $z$ is shown in the bottom-right panel of Figure 1.

 All the galaxies in this sample have \Sersic parameters
given by Blanton et al. (2002b). The stellar mass sample  by
Kauffmann et al. (2002a) contains objects spectroscopically
classified as galaxies, with magnitude in the range
$14.5<r<17.77$, selected from all available spectroscopic
observations in the SDSS upcoming Data Release One (DR1). The area
covered by this stellar mass sample is almost all contained within our
sample, and we obtain a subsample of 118,850 galaxies with stellar
masses.  This subsample has the same properties as the main sample
except the smaller sky coverage.

To study the size distribution of galaxies, we need to define a
complete sample for which selection effects can be corrected. As
discussed in Strauss et al. (2002), the stable $\target$ version
gives an almost complete sample in the magnitude range $15.0\leq
r\leq17.77$ and with surface-brightness $\mu_{50}\leq24.5\SB$.
However, during the commissioning of SDSS, a number of tentative
versions of $\target$ were used for refining the final $\target$
algorithms, and the trial $\target$s have small differences in the
magnitude and surface-brightness limits (see EDR for detail).
Therefore, to define a  complete sample we need to consider
selection effects in more detail. Because galaxies with
$23.0\SB<\mu_{50}<24.5\SB$ are targeted only when the local and
global sky values are within  0.05 $\SB$ (Strauss et al. 2002), we
set a lower surface-brightness  limit at $\mu_{\rm{lim}}=23.0\SB$.
As one can see from the bottom-left panel of Fig.~1, the total
number excluded by this selection criterion is very small. Next,
to avoid the contamination by bright stars,
$\target$ rejects bright compact objects with $R_{50}<2\acrsec$
and $r<15.0$ (15.5 in $\target$ v$\_2\_7$). Because of this, we
exclude all galaxies brighter than 15.0 (15.5 for objects targeted
by $\target$ v$\_2\_7$). As shown in Fig. 1, only a small number
of galaxies are excluded by this criterion also. Finally, the magnitude
limit ($r_{\rm max}$) at the faint end varies across the sky in
different versions of $\target$. We take this into account by
treating $r_{\rm max}$ as a function of sky position ($\theta$,
$\phi$).

A more important effect is that some galaxies are so small
(compact) that either their size measurements are seriously
affected by the PSFs, or they are misclassified as stars by
$\target$.  As discussed by Strauss et al. (2002), very  few
true galaxies at the compact end are missed by the target criteria.
However, to take care of the seeing effects, we use only galaxies with
angular sizes $R_{50}>D_{\rm{min}}$, and we choose
$D_{\rm{min}}=1.6\arcsec$ (i.e. 4 pixels).
This choice, based on the fact that the median seeing condition
in SDSS is about 1.5\arcsec, is  conservative, because
the PSF is known quite accurately. In practice, this cut
does not affect our results, as only a relatively small
fraction of galaxies is excluded (see the top-left panel of Fig.1).
Finally we also exclude a small number of galaxies with redshift
$z<0.005$, whose distances may be severely contaminated by their
peculiar velocities.
In summary,  our final complete sample includes all galaxies with
$\mu_{50}\leq23.0\SB$, $r_{\rm min}(\theta,\phi)\leq r\leq r_{\rm
max}(\theta,\phi)$, $R_{50}\geq1.6\arcsec$ and $z\geq0.005$. This
sample contains 138,521 galaxies, of which 99,786 have stellar
masses.

\subsection{Subsamples of galaxy types}

 In this paper we also wish to analyze the dependence
of the size distribution on galaxy type, so we need
to adopt some criteria to classify galaxies.

 There are attempts to classify SDSS galaxies into
morphological classes through direct inspection of the galaxy
images (Shimasaku et al. 2001; Nakamura et al. 2003). While such
eye-ball classification should match the original Hubble
morphological sequence, it is quite tedious and has so far been
carried out only for about 1500 big galaxies in the SDSS. However,
it has been suggested that some photometric and spectroscopic
properties may be closely correlated with the morphological type,
and so can be used as morphology indicators. For example,
Shimasaku et al. (2001) show that the concentration $c$ can be
used to separate early-type (E/S0) galaxies from late-type
(Sa/b/c, Irr) galaxies. Using about 1500 galaxies with eye-ball
classifications, Nakamura et al. (2003) confirmed that $c=2.86$
separates galaxies at S0/a with a completeness of about 0.82 for
both late and early types. For the \Sersic profile (Blanton et al.
2002b), the profile index $n$ is uniquely related to the
concentration parameter, and so the value of $n$ may also serve as
a morphological indicator. Other profile indicators of galaxy type
include the exponential and de Vaucouleurs profile likelihoods,
$P_{exp}$ and $P_{dev}$, given in the $\photo$ output. Based on
the broad band colors, Strateva et al. (2002) suggested that the
color criterion $u^*-r^*>2.22$ can separate early types (E/S0/Sa)
from late types (Sb/Sc/Irr). Blanton et al. (2002b) found that the
color  criteria $^{0.1}(g-r)\sim0.7$ [where $^{0.1}(g-r)$ is the
$g-r$ color K-corrected to the redshift of 0.1] separates galaxies
into two groups with distinct properties.  There are also
attempts to classify SDSS galaxies according to their  
spectral types, such as that based on   
the Principal Component Analysis (Yip et al. 2003) 
and that on the 4000$\rm{\AA}$ spectral break index  
(Kauffmann et al. 2002b). It must be pointed out, however,
that all these simple type classifications have uncertainties 
and are only valid in the statistical sense.  
For example, the profile and color indices can both be 
affected by dust extinction, while the
classifications based on spectra can be affected 
aperture biases due to the finite ($3\arcsec$
in diameter) of the fibers. Because of these 
uncertainties, we only divide galaxies into a small 
number of subsamples according to types.  
More specifically, we use $c=2.86$ and $n=2.5$ as two
basic indicators to separate galaxies into 
early and late types. With such a separation, most Sa
galaxies are included in the late-type category. We also use
the color criterion, $^{0.1}(g-r)=0.7$ for comparison. The $n$
separation is set at 2.5, the average between  exponential profile
($n=1$) and de Vaucouleurs profile ($n=4$), which also  gives an
early/late ratio similar to that given by the separator $c=2.86$.
We adopt the $^{0.1}(g-r)$ color rather than the $u^*-r^*$ color,
because the $g$-band photometry is currently better than the $u$-band
photometry in the SDSS and because the K-correction for the
$u$-band is very uncertain.

\section{The size distribution of galaxies}

In this section, we derive the size distribution as a function of
luminosity and stellar mass for galaxies of different types.
Specifically, we first bin galaxies of a given type into
small bins of absolute magnitude (or mass). We then use a $V_{\rm
max}$ method to make corrections for the incompleteness due to
selection effects, and derive the conditional size distribution
function $f_i(R|M_i)$ for a given bin. Finally, we investigate the
size distribution as a function of luminosity (or stellar mass).

\subsection{The $V_{\rm max}$ correction of the selection effects}

 As described in the last section, our sample is selected
to be complete only to some magnitude, size and surface-brightness
limits. In order to obtain the size distribution for the galaxy
population as a whole, we must make corrections for these selection
effects. In this paper we use the $V_{\rm max}$ method to do this.

The basic idea of the $V_{\rm max}$ method is to give each galaxy
a weight which is proportional to the inverse of the maximum
volume ($V_{\rm{max}}$) within which galaxies identical to the one
under consideration  can be observed. For a given galaxy with
magnitude $r$, Petrosian half-light radius $R_{50}$,
surface-brightness $\mu_{50}$, and redshift $z$, the selection
criteria described in the last section define the value of $V_{\rm
max}$ in the following way. First, the magnitude range $r_{\rm
min}\leq r\leq r_{\rm max} $ corresponds to a maximum redshift
$z_{\rm{max},m}$ and a minimum redshift $z_{\rm{min},m}$:
\begin{equation}
d_L(z_{\rm{max},m})=d_L(z)10^{-0.2(r-r_{\rm max})}\,;
~d_L(z_{\rm{min},m})=d_L(z)10^{-0.2(r-r_{\rm min})},
\end{equation}
where $d_L (z) $ is the luminosity distance at redshift $z$. Note
that we have neglected the effects of K-correction and luminosity
evolution in calculating $d_L(z_{\rm{max},m})$ and
$d_L(z_{\rm{min},m})$. In general, the K-correction make a given
galaxy fainter in the observed $r$-band if it is put at higher
redshift. The luminosity evolution has an opposite effect; it
makes galaxies brighter at higher redshift. We found that
including these two opposing effects (each is about 
one magnitude per unit redshift, see Blanton et al. 2002a;
2002c) has a negligible impact on our results.

 The surface-brightness limit constrains the $V_{\rm
max}$ of a galaxy mainly through the dimming effect. The maximum
redshift at which a galaxy of surface-brightness $\mu_{50}$ at $z$
can still be observed with the limit surface-brightness $\mu_{\rm
lim}=23.0$ is given by
\begin{equation}
z_{\rm{max},\mu}=(1+z)10^{{\frac{(23.0-\mu_{50})}{10}}}-1.
\end{equation}
Here, again, K-correction and luminosity evolution
are neglected. We have also neglected  possible color
gradients in individual galaxies.
The minimum size limit $D_{\rm{min}}$
also defines a maximum redshift $z_{\rm{max},R}$ given by
\begin{equation}\label{z_max,s}
\frac{d_A(z_{\rm{max},R})}{d_A(z)}=\frac{R_{50}}{1.6\acrsec},
\end{equation}
where $d_A$ is the
angular-diameter distance. The real maximum and minimum redshifts,
$z_{\rm{max}}$ and $z_{\rm{min}}$, for a given galaxy are
therefore given by
\begin{equation}
z_{\rm{min}}=\rm{max}(z_{\rm{min},m},0.005)\,;
~z_{\rm{max}}=\rm{min}(z_{\rm{max},m},z_{\rm{max},\mu},z_{\rm{max},R}),
\end{equation}
and the corresponding $V_{\rm max}$ is
\begin{equation}\label{V_max}
V_{\rm max}=\frac{1}{4\pi}\int d\Omega f(\theta,\phi)\int_{z_{\rm
min}(\theta,\phi)}^{z_{\rm
max}(\theta,\phi)}\frac{d_A^2(z)}{H(z)(1+z)}c\,dz\,,
\end{equation}
where $H(z)$ is the Hubble constant at redshift $z$, $c$ is
the speed of light, $f(\theta,\phi)$ is the sampling
fraction as a function
of position on the sky, and $\Omega$ is the solid angle.

The apparent-magnitude limit only influences the number of
galaxies at a given absolute-magnitude, and so it does not matter
when we analyze the size distribution for galaxies with a given
absolute magnitude. We can define a `conditional' maximum volume,
\begin{equation}{\label{c_Vmax}}
V^*_{\rm{max}}=\frac{V_{\rm{max}}} {(4\pi)^{-1}\int d\Omega
f(\theta,\phi)\int_{z_{\rm{min}}(\theta,\phi)}^{z_{\rm{max},m}(\theta,\phi)}
d_A^2(z) H^{-1}(z)(1+z)^{-1}c\,dz}\,
\end{equation}
which takes values from 0 to 1, and gives the probability a galaxy
with size $R_{50}$ can be observed at the given absolute
magnitude. Given $N$ galaxies in an absolute-magnitude (or mass)
bin $M\pm \Delta M$, the intrinsic conditional size distribution
$f(R|M)$ can be estimated from
\begin{equation}
f(R|M)\propto\sum_{i=1}^{N}\frac{1}{V^*_{{\rm max},i}}~~~
\mbox{if $R-dR<R_i<R+dR$},
\label{f(R|M)}
\end{equation}
where $R_i$ and $V^*_{{\rm{max}},i}$ are the radius and the value
of $V^*_{\rm max}$ for the $i$th galaxy.

\subsection{Size distribution: dependence on luminosity}

In this subsection, we study the size distribution as a function
of luminosity for galaxies of different type. The absolute magnitude $M$ is
calculated from the observed apparent
magnitude $m$ using
\begin{equation}\label{M}
M=m-DM(z)+5-K(z),
\end{equation}
where $z$ is the redshift of the galaxy, $DM(z)$ is the distance
modulus and $K(z)$ is the K-correction. The distances are
calculated from redshifts using a cosmology with mass density
$\Omega_0=0.3$, cosmological constant $\Omega_{\Lambda}=0.7$, and
Hubble's constant $h=0.7$.  The K-correction is calculated based
on the study of Blanton  et al. (2002a).

\begin{figure}
\epsfysize=14.0cm \centerline{\epsfbox{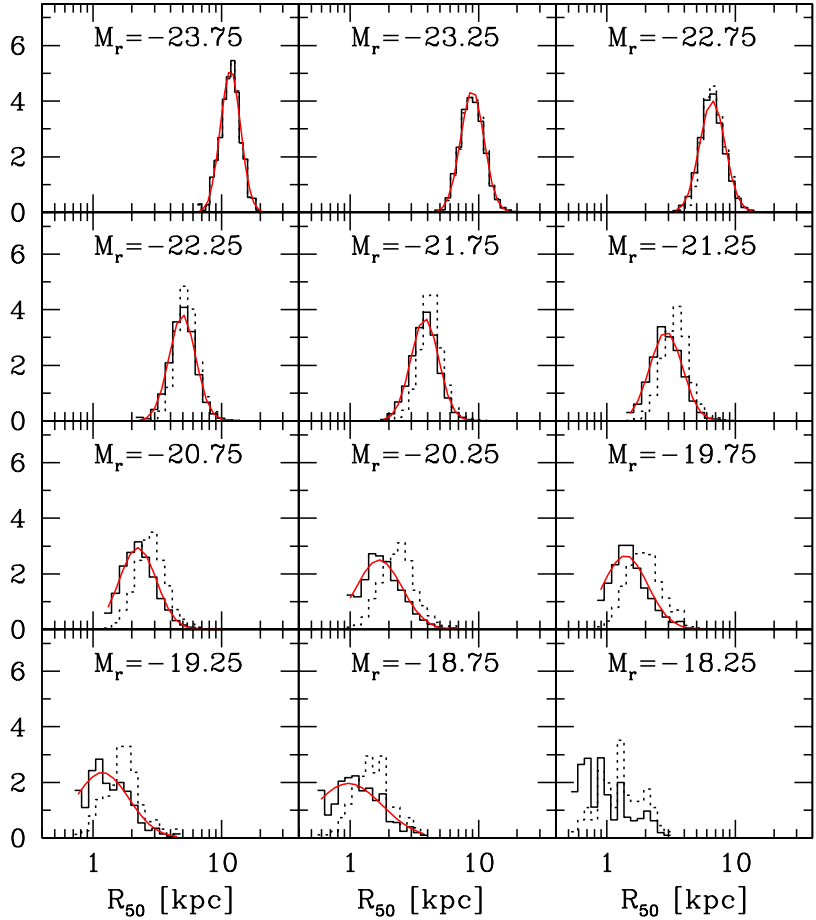}} \caption{
Histograms of Petrosian half-light radius $R_{50}$ (in the
$r$-band) for early-type ($c>2.86$) galaxies in different
Petrosian $r$-band absolute-magnitude bins. The dotted histograms
show the raw distribution, while the solid histograms show the
results after $V_{\rm{max}}$ correction for selection effects. The
solid curves are obtained by fitting the sizes to a log-normal
distribution through the maximum-likelihood method.}
\label{histEr}
\end{figure}

\begin{figure}
\epsfysize=14.0cm \centerline{\epsfbox{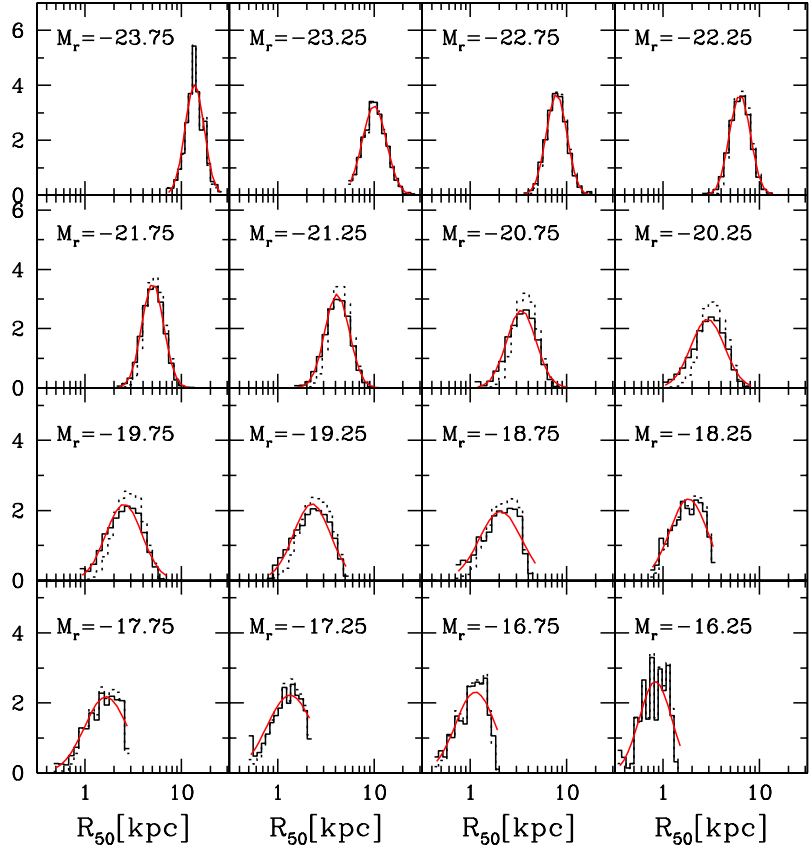}}
\caption{The same as Fig.\ref{histEr}, but
for late-type ($c<2.86$) galaxies.} \label{histLr}
\end{figure}

In Figures \ref{histEr} and \ref{histLr} we show the histograms of
Petrosian half-light radius $R_{50}$ for galaxies of different
absolute magnitudes and types. We use $c=2.86$ to separate
galaxies, in which case 32 percent of them are included in the early
types. Galaxies of a given type are further divided into
absolute-magnitude bins with a width of 0.5\,mag. The dotted
histograms show the observed size distributions, obtained by
directly counting the numbers of galaxies in given size bins. The
intrinsic distributions, obtained by using the $V^*_{\rm{max}}$
correction [see equation (\ref{f(R|M)})], are shown as the solid
histograms. All the histograms are normalized to the unit area in the space of
$\Log(R_{50})$.

As one can see, for both late- and early-type galaxies, the
intrinsic size distributions can be approximated
reasonably well by a log-normal
function. As we will see later, this type of distribution in
sizes is also motivated by theoretical considerations. We
therefore make the assumption that $f(R|M)$ has a log-normal form,
\begin{equation}
f(R,\bar{R}(M),\sigma_{\ln{R}}(M))=\frac{1}{\sqrt{2\pi}\sigma_{\ln{R}}(M)}
  \exp\left[-\frac{\ln^2(R/\bar{R}(M))}{2\sigma_{\ln{R}}^2(M)}\right]
  \frac{dR}{R},\label{lognormal}
\end{equation}
which is characterized by the median $\bar{R}(M)$ and the
dispersion $\sigma_{\ln{R}}(M)$.  We use a maximum likelihood
method to estimate $\bar{R}$ and $\sigma_{\ln{R}}$ at each
magnitude bin. The procedure goes as follows. For a sample of $N$
galaxies (in a certain absolute-magnitude bin) with sizes
$\{R_i\}_{i=1,N}$ and conditional maximum volumes
$\{V^*_{\rm{max},i}\}_{i=1,N}$, the likelihood for the size
distribution is
\begin{equation}
{\cal L}(\bar{R},\sigma_{\ln{R}})=
\prod_{i=1}^N\frac{1}{V^*_{\rm{max,i}}}
\frac{f(R_i,\bar{R},\sigma_{\ln{R}})dR} {\int_{R_{\rm{min}}}^
{R_{\rm{max}}}f(R,\bar{R},\sigma_{\ln{R}})dR},
\end{equation}
where $R_{\rm{min}}$ and $R_{\rm{max}}$ are the minimum and
maximum radii that can be observed for the luminosity bin in
consideration, $f(R,\bar{R},\sigma_{\ln{R}})$ is the log-normal
function with median $\bar{R}$ and dispersion $\sigma_{\ln{R}}$
given in equation (\ref{lognormal}). By maximizing this likelihood
function, we obtain the best estimates of $\bar{R}$ and
$\sigma_{\ln{R}}$ for each magnitude bin. The solid curves in
Figures 2 and 3 show the results of the log-normal functions so
obtained. As one can see, they provide very good fits to the solid
histograms.

\begin{figure}
\epsfysize=14.0cm \centerline{\epsfbox{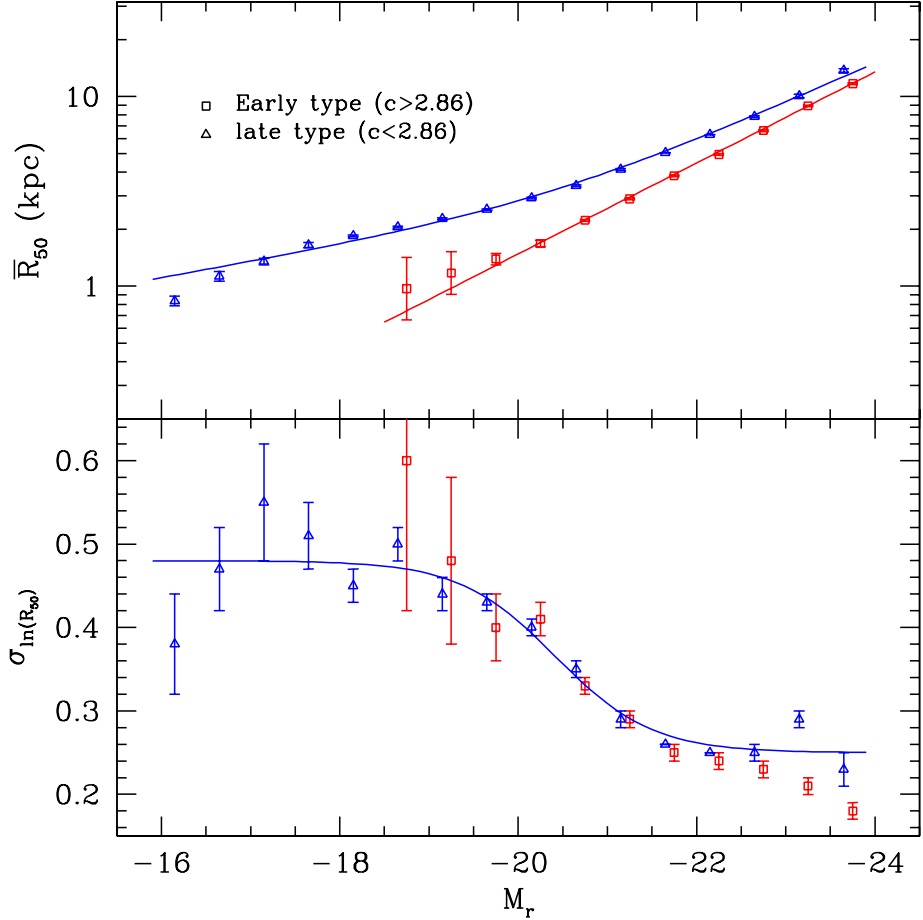}} \caption{The
median and dispersion of the distribution of Petrosian half-light
radius $R_{50}$ (in the $r$-band), as functions of $r$-band
Petrosian absolute magnitude, obtained by fitting a log-normal function.
Results for late-type ($c<2.86$) and early-type ($c>2.86$)
galaxies are shown as triangles and squares, respectively. The
error bars represent the scatter among 20 bootstrap samples. The
solid curves are the fit of the $\bar{R}$ - $M$ and
$\sigma_{\ln{R}}$ - $M$ relations by equations (\ref{fitERM}),
(\ref{fitLRM}) and (\ref{fitsigma}).} \label{RdisMr}
\end{figure}

Figure \ref{RdisMr} shows $\bar{R}$ (upper panel) and
$\sigma_{\ln{R}}$ (lower panel) against the absolute magnitude.
Triangles and squares denote the results for late- and early-type
galaxies, respectively. The error bars are obtained from the
scatter among 20 bootstrap samples. The small error bars show the
statistics one can get from the current sample. The number of
faint early-type galaxies is still too small to give any
meaningful results (see the panel of $M_r=18.25$ in
Fig. 1). This small number may not mean that the number of faint
elliptical galaxies is truly small; it may just reflect our
definition of early- and late-type galaxies. Indeed, faint
elliptical galaxies seem to have surface-brightness profiles better
described by an exponential than a $R^{1/4}$ law (Andredakis,
Peletier \& Balcells 1995; Kormendy \& Bender 1996), and so they
will be classified as `late-type' galaxies according to the $c$
criterion because of their small concentration.

As shown in Fig. 4, the dependence of $\bar{R}$ on the absolute
magnitude is quite different for early- and late-type galaxies. In
general, the increase of $\bar{R}$ with luminosity is faster for
early-type galaxies. The $\bar{R}$ - $M$ relation can roughly be
described by a single power law for bright early-type galaxies,
while for late-type galaxies, the relation is significantly curved, with
brighter galaxies showing a faster increase of $\bar{R}$ with
$M$. In the luminosity range where $\bar{R}$ and $\sigma_{\ln R}$
can be determined reliably, the dispersion has a similar trend with
$M$ for both early- and late-type galaxies. An interesting feature
in $\sigma_{\ln R}$ is that it is significantly smaller for
galaxies brighter than $-20.5$ mag (in the $r$-band). As we will
discuss in Section 4, these observational results have important
implications for the theory of galaxy formation.

To quantify the observed $\bar{R}$ - $M$ and $\sigma_{\ln R}$ -
$M$ relations, we fit them with simple analytic
formulae. For early-type galaxies, we fit $\bar{R}$ - $M$ by,
\begin{equation}\label{fitERM}
\Log(\bar{R}/\kpc)=-0.4aM+b\,,
\end{equation}
where $a$ and $b$ are two fitting constants.
For late-type galaxies, we fit the size-luminosity
relation and its dispersion by
\begin{equation}\label{fitLRM}
\Log(\bar{R}/\kpc)=-0.4\alpha M+
(\beta-\alpha)\Log[1+10^{-0.4(M-M_0)}]+\gamma
\end{equation}
and
\begin{equation}\label{fitsigma}
\sigma_{\ln R}=\sigma_2
+\frac{(\sigma_1-\sigma_2)}{1+10^{-0.8(M-M_0)}},
\end{equation}
where $\alpha$, $\beta$, $\gamma$, $\sigma_1$, $\sigma_2$ and
$M_0$ are fitting parameters. Note that the value of $M_0$ used in
equation (\ref{fitLRM}) is determined by fitting the observed
$\sigma_{\ln R}$ - $M$ relation [equation (\ref{fitsigma})],
because the fit of the $\bar{R}$ - $M$ relation is not very
sensitive to the value of $M_0$. Thus, the relation between
$\bar{R}$ and the luminosity $L$ is $R\propto L^{a}$ for
early-type galaxies. For late-type galaxies, $R\propto L^{\alpha},
\sigma_{\ln R}=\sigma_1$ at the faint end ($L\ll L_0$, where $L_0$
is the luminosity corresponding to $M_0$), and $R\propto
L^{\beta}, \sigma_{\ln R}=\sigma_2$ at the bright end ($L\gg
L_0$).  We use the least-square method to estimate the fitting
parameters and the results are given in Table 1. These fit results
are also plotted as solid lines in  Fig. 4.

\begin{figure}
\epsfysize=14.0cm \centerline{\epsfbox{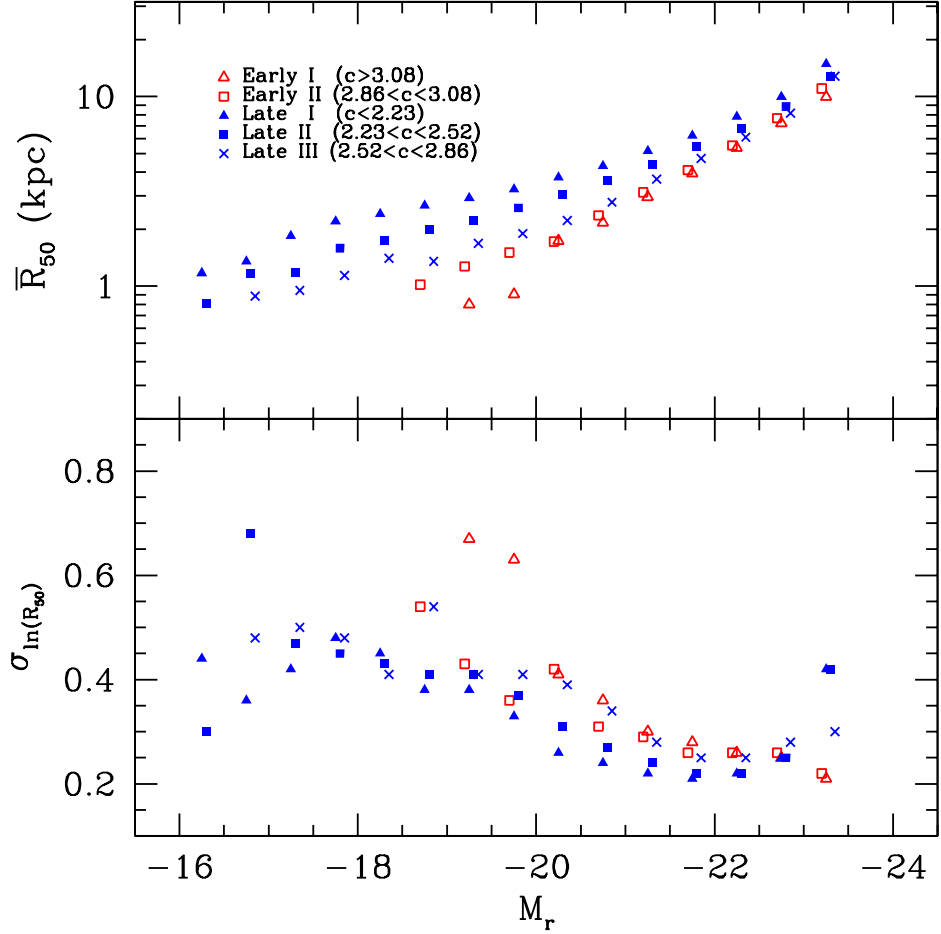}} \caption{The
median and dispersion of the $R_{50}$ (in the $r$-band)
distribution, as a function of $r$-band Petrosian absolute magnitude, for
galaxies in fine bins of $c$.} \label{RdisMcr}
\end{figure}

We have also analyzed the size distribution as a function of
luminosity in finer ranges of $c$. Specifically, we further divide
late-type galaxies ($c<2.86$) into three sub-samples containing
equal numbers of galaxies, and early-type galaxies ($c>2.86$) into
two equal sub-samples. The ranges of $c$ and the results for
$\bar{R}$ and $\sigma_{\ln R}$ for these sub-samples are shown in
Figure 5. As we can see, the $\bar{R}$ - $M$ relation depends
systematically on $c$: galaxies with higher $c$ show a steeper
relation. However, the difference between the two
early-type samples is quite small, except for the
two faintest bins where the statistic is quite poor.

\begin{figure}
\epsfysize=14.0cm \centerline{\epsfbox{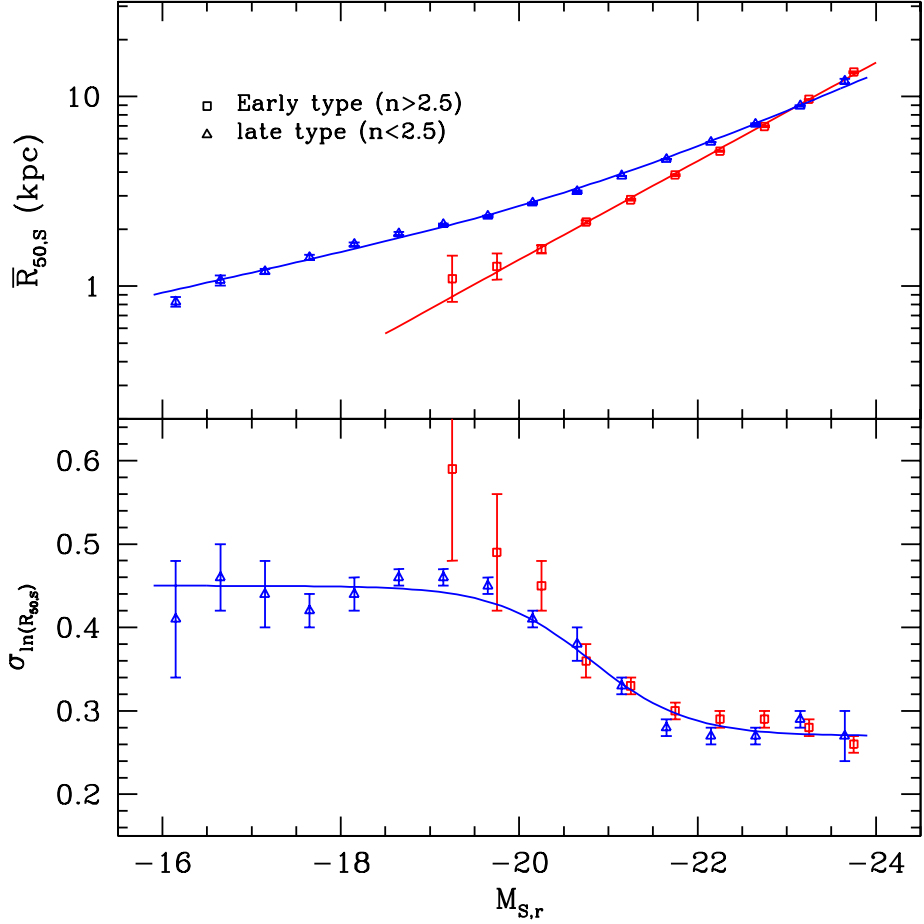}} \caption{The
median and dispersion of the distribution of the \Sersic
half-light radius $R_{50,S}$ (in the $r$-band) as functions of
$r$-band absolute \Sersic magnitude. Here a galaxy is separated into
early- or late-type according to whether its \Sersic index $n$ is
larger or smaller than $2.5$. The error bars represent the scatter
among 20 bootstrap samples. The solid curves are the fit of the
$\bar{R}$ - $M$ and $\sigma_{\ln{R}}$ - $M$ relations by equations
(\ref{fitERM}), (\ref{fitLRM}) and (\ref{fitsigma}).}
\label{SRdisMr}
\end{figure}

As discussed in Section 2.2, the \Sersic half-light radii
 $R_{50,S}$ given by Blanton et al. (2002b) have the merit of being
corrected for PSF and, unlike the Petrosian magnitude,  
the \Sersic magnitude has also the merit of including 
the total flux of a galaxy. We therefore also made analyses 
based on the \Sersic quantities. Here we use
$n=2.5$ to separate late- and early-type galaxies. In this
separation, about 36 percent of the galaxies are classified as
early types. Figure 6 shows the results of $\bar{R}$ and
$\sigma_{\ln R}$ for the \Sersic quantity $R_{50, S}$. 
We have also fitted the
size-luminosity relations to the functional form given by
equations (\ref{fitERM}), (\ref{fitLRM}) and (\ref{fitsigma}), and
the fitting parameters are  listed in Table 1.  
Comparing these results with those shown in Fig. 4,  
we see that early type galaxies here have  systemtical 
bigger half-light radii for given luminosity.
This is caused by differences between the Petrosian and 
\Sersic quantities. For a galaxy with pure de Vaucouleurs 
profile, the Petrosian magntitude includes about 80 percent 
of the total flux, while the Petrosian half-light radius 
is only about 70 percent of the real half-light radius. 
Note that our derived slopes for the $R-L$ relation 
in both cases are consitent with the result $R\propto L^{0.63}$ 
obtained by Bernardi et al. (2003b) where yet another 
photometric system is used. For late type galaxies with exponential light profiles,
no significant difference is found between these these 
two systems, because the Petrosian magnitude includes almost 
all the total flux and the half light radius is approximately the same as the true half light radius.
Although our results show a significant curved  $R-L$ relation for late type galaxies, 
a simple power law is usually used as an assumption in previous studies due to the small samples.
But the results are generally consistent, for example, the relation gotten by de Jong \& Lacey(2000) 
is $R\propto L^{0.25}$ in I band.

\begin{figure}
\epsfysize=14.0cm \centerline{\epsfbox{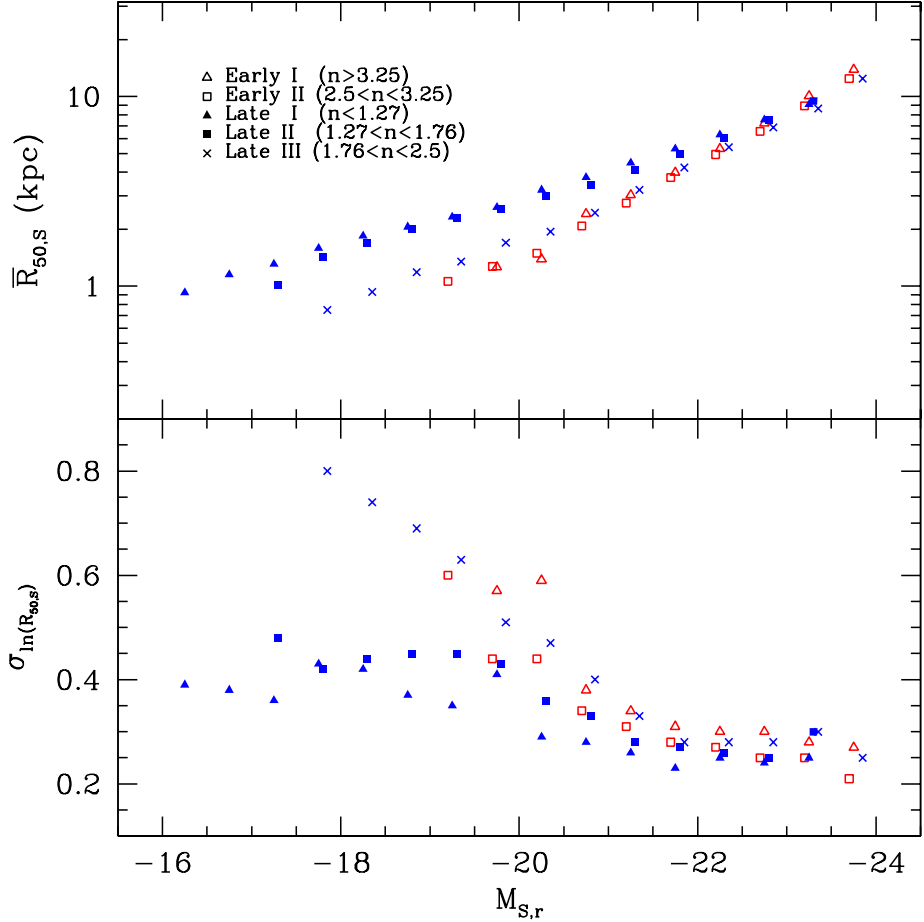}} \caption{The
median and dispersion of the distributions of \Sersic half-light
radius, $R_{50,S}$, as functions of $r$-band absolute \Sersic
magnitude, for galaxies in fine bins of the \Sersic index $n$.}
\label{SRdisMnr}
\end{figure}

Expanding on this, we have analyzed the size distribution as a function of
luminosity in finer bins of $n$. We divide
the late-type sample ($n<2.5$) into three equal sub-samples,
and the early-type sample ($n>2.5$) into two
equal sub-samples. The ranges of $n$ for these
sub-samples and the fitting results are shown in Figure 7. These
results should be compared with those shown in Fig. 5. While
galaxies with higher $n$ do show a steeper $\bar{R}$ - $M$
relation, the change of the trend with $n$ is less systematic than
with $c$. It seems that galaxies are separated into two
groups at $n\sim 1.7$, and galaxies in each group have similar
$\bar{R}$ - $M$ relations, independent of $n$. As shown by the two
dimensional distribution of galaxies in the space spanned by $n$
and the $^{0.1}(g-r)$ color (Blanton et al. 2002b), the cut at
$n=1.7$ roughly corresponds to a color cut at $^{0.1}(g-r)\approx
0.7$. The latter cut appears to separate E/S0/Sa
from Sb/Sc/Irr galaxies, as discussed in subsection 2.4.

\begin{figure}
\epsfysize=14.0cm \centerline{\epsfbox{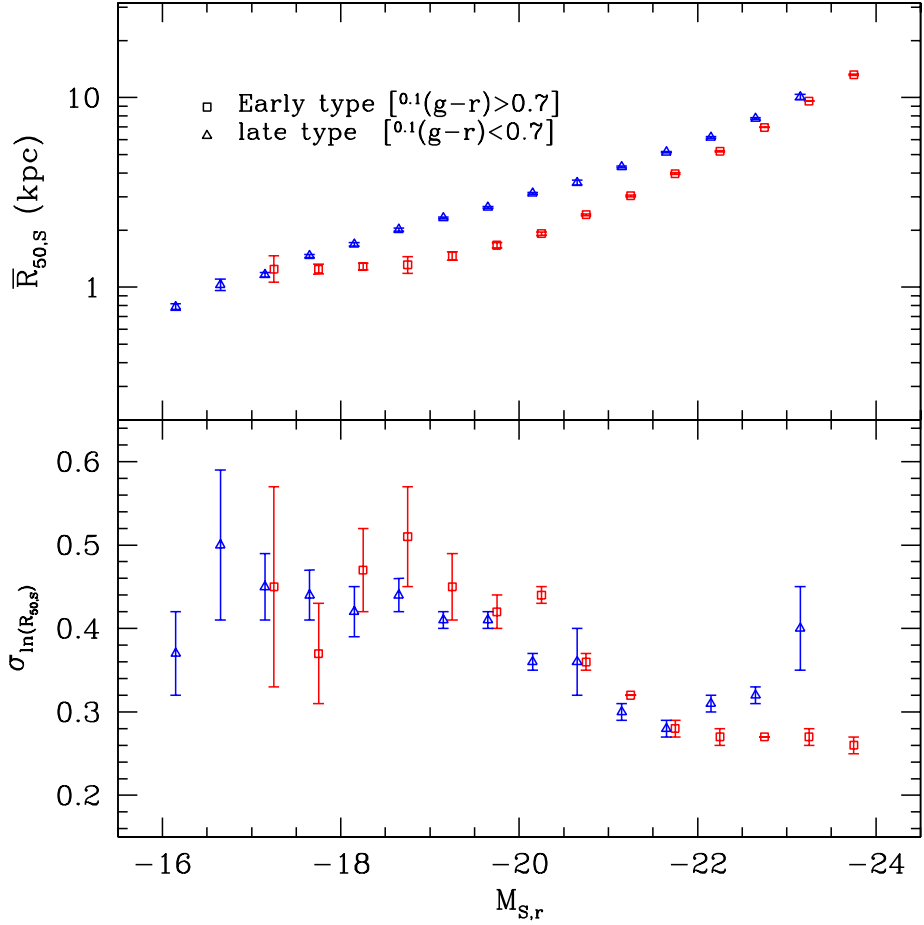}} \caption{The
median and dispersion of the distribution of \Sersic half-light
radius $R_{50,S}$, as a function of $r$-band \Sersic absolute magnitude.
Triangles represent results for late-type galaxies [here defined
to be those with $^{0.1}(g-r)<0.7$], while the squares are for
early-type galaxies with $^{0.1}(g-r)>0.7$. The error bars
represent the scatter among 20 bootstrap samples.}
\label{SRdisMcol}
\end{figure}

For comparison, we consider separating galaxies
according to the color criteria $^{0.1}(g-r)=0.7$. The results are
shown in Figure 8. In this case, since most Sa galaxies are
classified as early-type galaxies, there are fewer bright
late-type galaxies. Moreover, we begin to see faint red galaxies
(presumably faint ellipticals), which would be classified as
`late-type' galaxies by the $c$ and $n$ criteria, because of their
low concentrations. As one can see, the late-type galaxies show
approximately the same statistical properties as those in the $c$
and $n$ classifications. This is also true for bright early-type
galaxies. Red galaxies with $M_r\sim -20$ seem to follow a parallel
trend to
late-type galaxies, although they are smaller at given absolute
magnitude. This is consistent with the fact that many dwarf
ellipticals show exponential surface-brightness profiles, have
small sizes, and have size-luminosity scaling relations similar to
that of spiral galaxies (e.g. Caon, Capaccioli \& D'Onofrio 1993;
Kormendy \& Bender 1996; Guzman et al. 1997; Prugniel \& Simien 1997;
 Gavazzi et al. 2001). Faint red galaxies with  $M_r>
-19$ seem to have almost constant size. Note that $\sigma_{\ln{R}}$ -
$M$ show similar dependence on $M$ for both red and blue
galaxies.

\begin{figure}
\epsfysize=14.0cm \centerline{\epsfbox{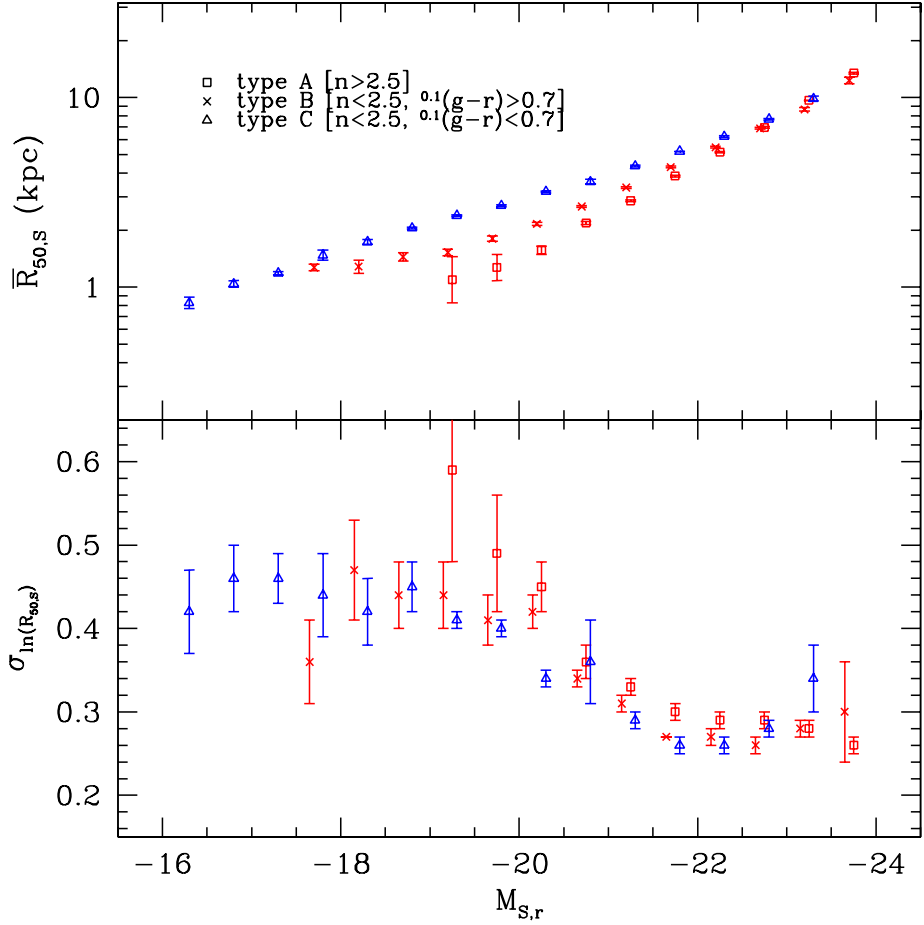}} \caption{The
median and dispersion of the distribution of \Sersic half-light
radius $R_{50,S}$ in $r$-band as a function of $r$-band absolute
\Sersic magnitude. The galaxies are separated into three subsamples
according to the \Sersic index $n$ and color $^{0.1}(g-r)$. The
error bars represent the scatter among 20 bootstrap samples.}
\label{SRMngr}
\end{figure}

Since most of the concentrated galaxies have red colors while
galaxies with low concentrations can have both blue and red
colors, it is interesting to examine the properties of galaxies
selected by both color and $n$. To do this, we consider a case
where galaxies with $n<2.5$ are divided further into two
subsamples according to the color criterion $^{0.1}(g-r)=0.7$. The
results are shown in Figure \ref{SRMngr}. As we can see,  B type
galaxies (with low $n$ and red color) show a $R$-$M$ relation
which is closer to that of A type (high $n$) galaxies than that of
$C$-type (blue and low $n$). Note again that faint red galaxies
have sizes almost independent of luminosity. The $\sigma_{\ln{R}}$
- $M$ relations are similar for all three cases.

\begin{figure}
\epsfysize=14.0cm \centerline{\epsfbox{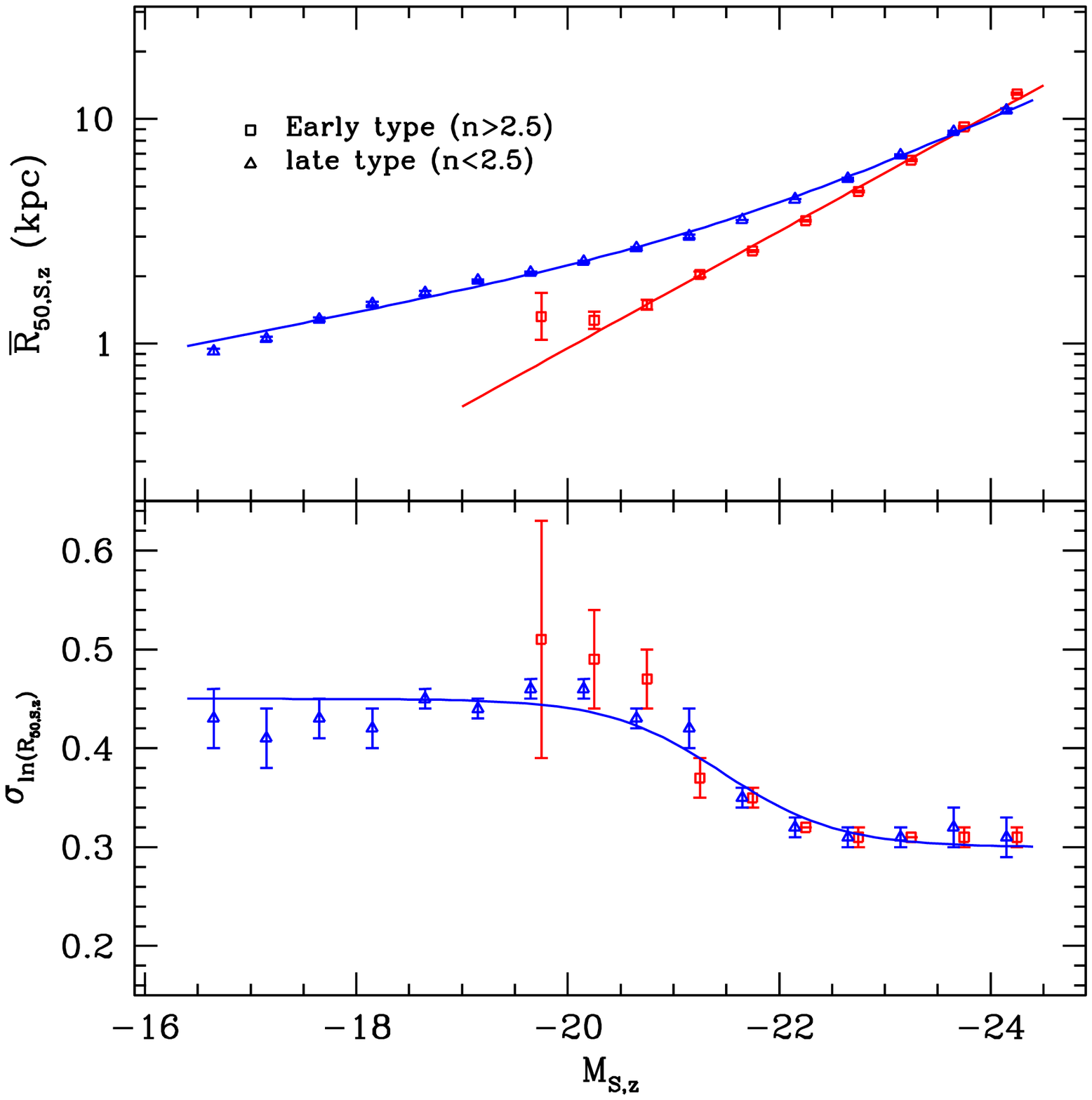}} \caption{The
median and dispersion of the distribution of \Sersic half-light
radius $R_{50,S}$ in the $z$-band, as a function of $z$-band
\Sersic absolute magnitude. Triangles represent results for late-type
galaxies (here defined to be those with $n<2.5$), while the
squares are for early-type galaxies (with $n>2.5$). The error bars
represent the scatter among 20 bootstrap samples. The solid curves
are the fit of the $\bar{R}$ - $M$ and $\sigma_{\ln{R}}$ - $M$
relations by equations (\ref{fitERM}), (\ref{fitLRM}) and
(\ref{fitsigma}).}
\label{SRdisMz}
\end{figure}

 So far our discussion has been based on the $r$-band data.
If galaxies possess significant radial color gradients, the size
of a galaxy may be different in different wavebands. Furthermore,
if galaxies have different colors, the size distribution as a
function of luminosity may also be different in different
wavebands. To test how significant these effects are, we have
analyzed the size distributions separately in the SDSS $g$, $i$
and $z$ bands, using either the absolute magnitudes in the
corresponding band or the absolute magnitudes in the $r$-band to
bin galaxies into luminosity sub-samples. The results are
qualitatively the same as derived from the $r$-band data.
Similar conclusions for early type galaxies have been 
reached by Bernardi et al. (2003b). As an example, we show in Figure
\ref{SRdisMz} the results based on \Sersic radii and \Sersic 
$z$-band magnitudes. The galaxies are also separated into late-
and early- type by the $r$-band \Sersic index $n=2.5$.  The
results of fitting the $\bar{R}$ - $M$ and $\sigma_{\ln R}$ - $M$
relations are presented in Table 1. Because of the long wavelength
involved in the $z$-band photometry, the quantities in this band
may better reflect the properties of the stellar mass (e.g.
Kauffmann et al. 2002b).

\begin{table}
\caption{The least square fitting results of the parameters in the
$\bar{R}$ - $M$ and $\sigma_{\ln{R}}$ - $M$ relations. Cases of
figure 4, 6 and 10 use the fitting formula in equations
(\ref{fitERM}), (\ref{fitLRM}) and (\ref{fitsigma}), while
equations (\ref{fitERMass}), (\ref{fitLRMass}) and
(\ref{fitsigMass}) are used for the case of Figure 11.}
\begin{tabular}{lcccccccccc} \hline
 &\multicolumn{2}{c}{Early type} & &\multicolumn{4}{c}{Late type} &
 &\multicolumn{2}{c}{Scatter}\\
\cline{2-3}\cline{5-8}\cline{10-11}
Case & $a$ & $b$ & & $\alpha$ & $\beta$ & $\gamma$ &$M_0$ & &$\sigma_1$
&$\sigma_2$ \\
\hline
 Figure 4 & 0.60 & $-4.63$ & & 0.21 & 0.53 & $-1.31$ & $-20.52$ & &0.48
 &0.25 \\
 Figure 6 & 0.65 & $-5.06$ & & 0.26 & 0.51 & $-1.71$ & $-20.91$ & &0.45
 &0.27 \\
 Figure 10 & 0.65 & $-5.22$ & & 0.23 & 0.53 & $-1.53$ & $-21.57$ & &0.45
 & 0.30\\
 Figure 11& 0.56 & $3.47\times10^{-5}$ & & 0.14 & 0.39 & 0.10
 & $3.98\times10^{10}\Msun$& &0.47 &0.34\\
\hline
\end{tabular}
\end{table}

\subsection{Size distribution: dependence on stellar mass}

\begin{figure}
\epsfysize=14.0cm \centerline{\epsfbox{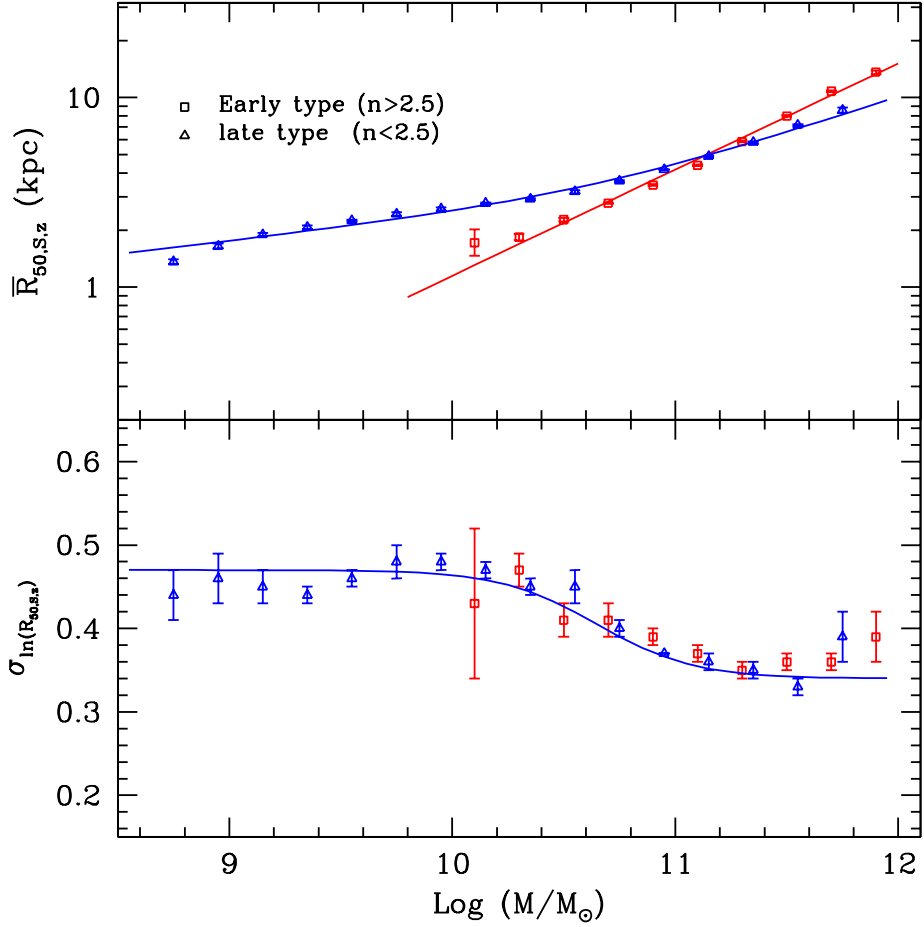}}
\caption{The median and dispersion of the distribution of \Sersic
half-light radius $R_{50,S}$ in the $z$-band, as a function of
stellar mass. Triangles represent results for late-type galaxies
(here defined to be those with $n<2.5$), while the squares are for
early-type galaxies (with $n>2.5$). The error bars represent the
scatter among 20 bootstrap samples. The solid curves are the fit
of the $\bar{R}$ - $M$ and $\sigma_{\ln{R}}$ - $M$ relations by
equations (\ref{fitERMass}), (\ref{fitLRMass}) and
(\ref{fitsigMass}).}
 \label{SRdisMassz}
\end{figure}

In this subsection, we study the size distribution of galaxies as
a function of stellar mass. We use the data obtained by Kauffmann
et al. (2002a). Figure \ref{SRdisMassz} shows the results based on
the $z$-band \Sersic half-light radii.  To quantify the mass
dependence of $\bar{R}$ and $\sigma_{\ln R}$, we fit $\bar{R}$ -
$M$ relation for the early-type galaxies by
\begin{equation}\label{fitERMass}
\bar{R}(\kpc)=b\left(\frac{M}{\Msun}\right)^{a}\,.
\end{equation}
For late-type galaxies, we fit $\bar{R}$ - $M$ and $\sigma_{\ln
R}$ - $M$ by
\begin{equation}\label{fitLRMass}
\bar{R}(\kpc)=\gamma\left(\frac{M}{\Msun}\right)^{\alpha}
\left(1+\frac{M}{M_0}\right)^{\beta-\alpha}
\end{equation}
and
\begin{equation}\label{fitsigMass}
\sigma_{\ln R}=\sigma_2+\frac{(\sigma_1-\sigma_2)}{1+(M/M_0)^2}\,,
\label{sigmamass}
\end{equation}
respectively,
where $M$ is the stellar mass, $\alpha$, $\beta$, $\gamma$,
$\sigma_1$, $\sigma_2$, $M_0$, $a$ and $b$ are all fitting
parameters.  Those parameters have the same meaning as those in equations
 (\ref{fitERM}), (\ref{fitLRM}) and (\ref{fitsigma}) except
that stellar mass is used instead of luminosity.  The values of
these parameters given by a least square fitting to the data are
also listed in Table 1. The fitting results are shown as the solid
curves in Fig.  \ref{SRdisMassz}. Here, similar to the
size-luminosity relation, $M_0$ is the characteristic mass at
which $\sigma_{\ln R}$ changes significantly and  is about
$10^{10.6}\Msun$. For late type galaxies, the low-mass galaxies
($M\ll M_0$) have $\bar{R}\propto M^{0.14}$ and $\sigma_{\ln
R}=0.47$, and the high-mass galaxies ($M\gg M_0$) have
$\bar{R}\propto M^{0.39}$ and $\sigma_{\ln R}=0.34$. The early
type galaxies follow the relation $\bar{R}\propto M^{0.56}$. The
power indices $a$, $\alpha$ and $\beta$ are smaller for the mass
than for the luminosity, because the mass-to-light ratio is
systematically higher for galaxies with higher luminosity.

\begin{figure}
\epsfysize=14.0cm \centerline{\epsfbox{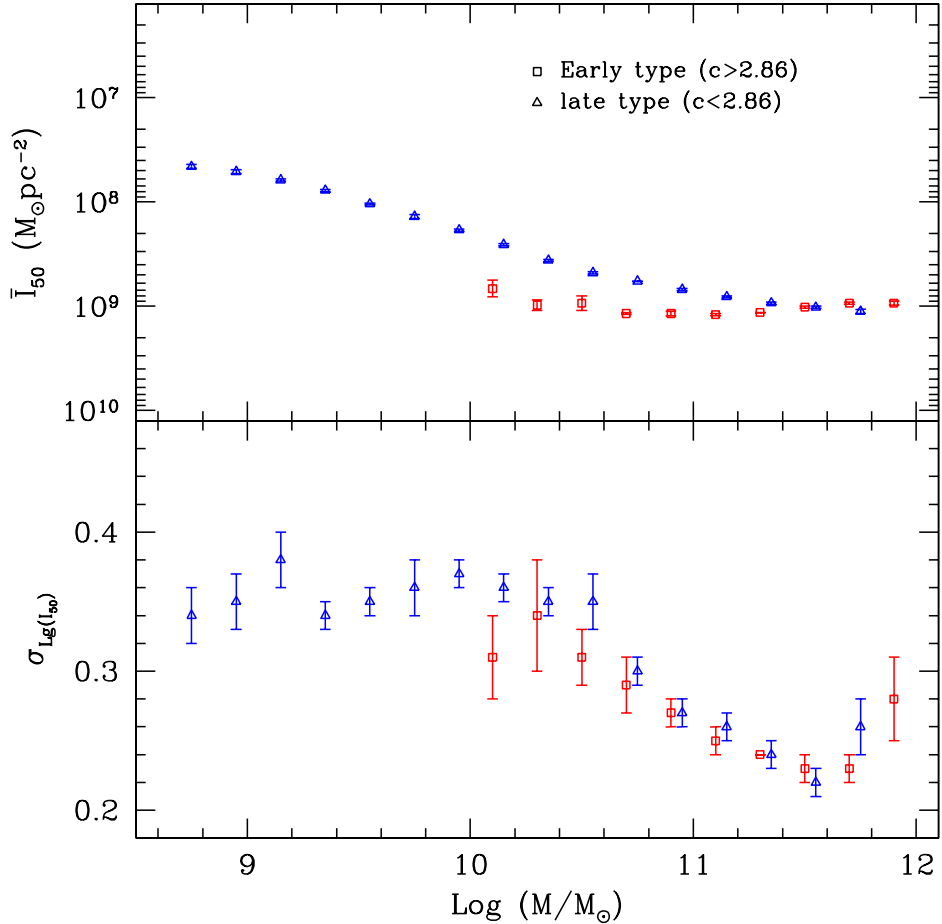}} \caption{The
distributions of the average surface-mass density $I_{50}$ within
$R_{50,z}$ as functions of stellar masses. Error bars represent
scatter among 20 bootstrap samples.} \label{SDdisMass}
\end{figure}

For early-type galaxies, the power index $a=0.56$ implies that
the average surface mass density $I_{50}$ within the half-light radius is
roughly a constant, which is shown directly in Fig.
\ref{SDdisMass}. Here we have assumed  that no mass-to-light
ratio gradient exists in the $z$-band so that the half-light
radius also encloses half of the stellar mass.
Moreover, since the stellar mass is derived by multiplying the Petrosian luminosity
with the model derived mass-to-light ratio $M/L$ (Kauffmann et al., 2002a),  the 
Petrosian half light radius $R_{50}$ is used here in calculating the $I_{50}$.
As we will discuss in Section 5, these results have important
implications for the formation of elliptical galaxies.

\subsection{The surface-brightness distribution}

\begin{figure}
\epsfysize=14.0cm \centerline{\epsfbox{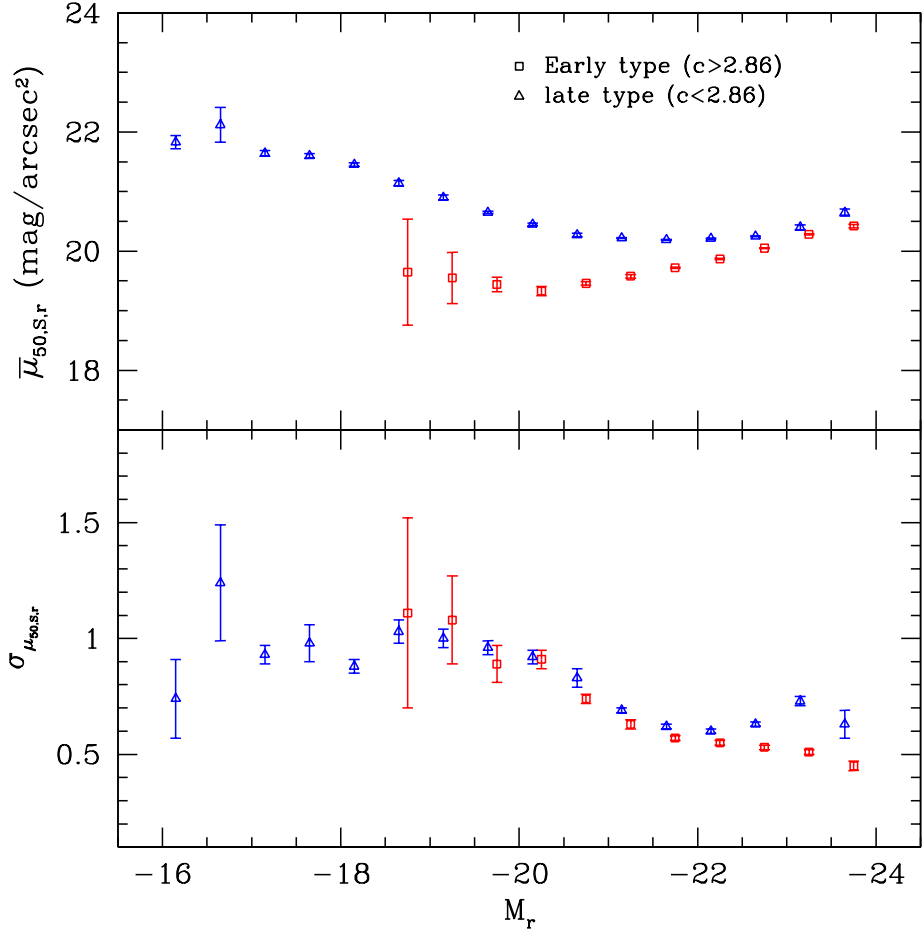}} \caption{The
median and dispersion of the distribution of $r$-band effective
surface brightness(defined in \Sersic system) as functions of 
$r$-band Petrosian absolute magnitude.
Triangles represent results for late-type galaxies (here defined
to be those with $c<2.86$), while the squares are for early-type
galaxies (with $c>2.86$). The error bars represent the scatter
among 20 bootstrap samples. The two vertical lines denote the
observational surface brightness limits (see text). } \label{udisMr}
\end{figure}

The intrinsic surface-brightness $\mu_{50}$  of a galaxy is linked
to its size $R_{50}$ through
\begin{equation}\label{Rtomu}
\mu_{50}(\SB)=M(\rm{mag})+5\Log\,R_{50}(\kpc)+38.57,
\end{equation}
where $M$ represent the absolute magnitude. Thus, the log-normal
size distribution for a given luminosity implies that the
surface-brightness distribution at a given luminosity is normal.
In this case, we may obtain the median and dispersion of the
surface-brightness distribution directly from the size
distribution through equation (\ref{Rtomu}). However, since the
width of our magnitude bins (0.5mag) is finite, the conversion is
not accurate. We therefore recalculated the median and dispersion
of the surface-brightness distribution in each magnitude bin by
using the same maximum-likelihood method as for the size
distribution. As an example, we show the $r$-band
surface-brightness distribution in Figure \ref{udisMr}.
The surface-brightness used here is defined in the \Sersic
system, i.e. the average surface brightness inside the \Sersic
half-light radius, but galaxies are labelled by their 
Petrosian magnitudes. The reason for this is that the 
luminosity function we are going to use to derive the 
integrated surface brightness distribution is based on Petrosian 
magnitudes. As before, galaxies are separated into 
late- and early-type galaxies at
$c=2.86$. As one can see from Fig. \ref{udisMr},
the brighter late-type galaxies have
systematically higher surface-brightness while the trend is the
opposite for bright early-type galaxies. This is the well-known
Kormendy relation (1977). Another feature clearly seen is that the
mean value of the surface-brightness is almost independent of
luminosity for bright late-type galaxies, which is consistent with
the Freeman disk (Freeman 1970). For dwarf late-type galaxies, the
surface brightness shows a strong increase with increasing of
luminosity in the range $-20<M_r<-18$. But the median value of the
surface brightness is consistent with being constant in the
luminosity range $-16<M_r<-18$. However, this result should be
treated with caution, because the median value is already quite
close to the limit $23.0 \SB$. Any incompleteness near $23.0 \SB$
can bias the median to a lower value (i.e. higher surface
brightness).

\begin{figure}
\epsfysize=14.0cm \centerline{\epsfbox{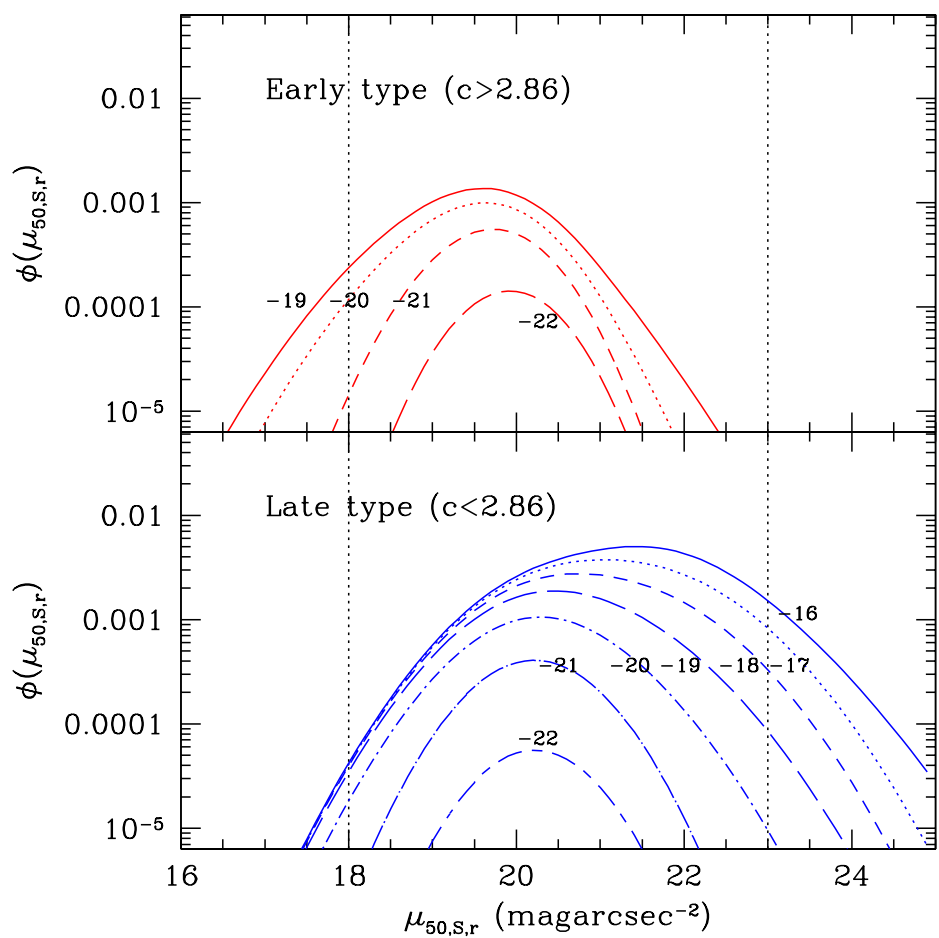}}
\caption{The surface-brightness distribution for late-
($c<2.86$) and early-type ($c>2.86$) galaxies in different
luminosity ranges, obtained by convolving the observed
luminosity function $\phi(L)$ with the conditional surface-brightness
distribution $f(\mu_{50}\vert L)$ shown in Fig. \ref{udisMr}.}
\label{IudisM}
\end{figure}

With the conditional surface-brightness distribution function
$f(\mu_{50}|M)$, we can calculate
the number density of galaxies at any given surface-brightness
$\mu_{50}$ by integrating over the luminosity function $\phi(M)$:
\begin{equation}
\phi(\mu_{50})=\int\phi(M)f(\mu_{50}|M)\,dM\,. \label{Iu}
\end{equation}
The luminosity functions of early- and late-type galaxies
separated at $c=2.86$ have recently been given 
by Nakamura et al. (2003) based on the Petrosian magnitudes.
As shown in Fig. \ref{udisMr}, our
conditional surface-brightness distribution is reliably determined
only in the luminosity range $-24<M_r<-16$ for late-type galaxies,
and in $-24<M_r<-19$ for early-type galaxies. Therefore, we set
the bright end of the integration in equation (\ref{Iu}) to be
$M_r=-24$ and carry out the integration from a number of
low-luminosity limits. The median and dispersion of the
surface-brightness distribution at any given magnitude are
obtained from a linear interpolation between adjacency magnitude
bins. The results are shown on Figure \ref{IudisM} with the
faint-end limits labelled on the corresponding curves. The two
vertical lines at $18.0$ and $23.0\SB$ correspond to the
observational surface brightness limits(defined in Petrosian system)
of our sample. The bright limit $18.0\SB$
corresponds to the brightest galaxies ($r=15.0$) with sizes at the
lower limit ($1.6\arcsec$).

As one can see from the figure, there may be many compact
early-type galaxies that are not included in our sample. For
late-type galaxies,  the surface-brightness shows a narrow normal
distribution for bright galaxies (i.e. the Freeman disk). When
more dwarf galaxies are included, low surface-brightness galaxies
may contribute a large fraction of the total numbers of galaxies.
Unfortunately, the current data cannot give a stringent constraint
on the number density of low-surface brightness galaxies because
the results for very faint galaxies ($M_r>-16$) are uncertain.
However, the fact that galaxies with the lowest surface brightness
are predominantly of low luminosity suggests that such low-surface
brightness galaxies contribute little to the luminosity density of
the universe. A similar conclusion has been reached in earlier
analysis (e.g. de Jong \& Lacey 2000; Cross \& Driver 2002;
Blanton et al. 2001).

\section{Theoretical expectations}

  In the preceding sections we have seen that the current
SDSS data can be used to derive good statistics for the
size distribution of galaxies and its dependence on luminosity,
stellar mass, concentration and color. In this section, we examine
whether or not these observational results can be accommodated in
the current paradigm of galaxy formation.

\subsection{Late -- type galaxies}

 Let us start with late-type (spiral) galaxies. A spiral galaxy
generally consists of  a rotationally supported thin disk,
and an ellipsoidal bu	lge which rotates relatively slowly.

\subsubsection {The disk component}

According to current theory of galaxy formation,
galaxy disks are formed as gas with some initial angular
momentum cools and contracts in dark matter haloes.
Our model of disk formation follows that described in
Mo, Mao \& White (1998 hereafter MMW). The model
assumes spherical dark haloes with density profile given
by  Navarro, Frenk \& White (1997, hereafter NFW):
\begin{equation}
 \rho(r)=\frac{\rho_0}
  {(r/r_s)(1+r/r_s)^2}\,,
\end{equation}
where $r_s$ is a characteristic radius, and $\rho_0$ is a
characteristic density. The halo radius $r_{200}$ is defined
so that the mean density within it is 200 times the critical
density. It is then easy to show that $r_{200}$ is related to the
halo mass $M_h$ by
\begin{equation}\label{r200}
 r_{200}={G^{1/3} M_h^{1/3}\over
[10 H(z)]^{2/3}}\,,
\end{equation}
where $H(z)$ is the  is the Hubble constant at redshift $z$.
The total angular momentum of a halo, $J$, is usually
written in terms of the spin parameter,
 \begin{equation}\label{lambda}
 \lambda=J|E|^{1/2}G^{-1}M_h^{-5/2},
\end{equation}
where $E$ is the total energy of the halo.
$N$-body simulations show that the distribution of halo
spin parameter $\lambda$ is approximately
log-normal,
 \begin{equation}\label{p_of_lambda}
p(\lambda)\,d\lambda=\frac{1}{\sqrt{2\pi}\sigma_{\ln\lambda}}
  \exp\left[-\frac{\ln^2(\lambda/\bar{\lambda})}
{2\sigma_{\ln\lambda}^2}\right]
  \frac{d\lambda}{\lambda},
\end{equation}
with $\bar{\lambda}\sim 0.04$ and $\sigma_{\ln\lambda}\approx0.5$
(Warren et al. 1992; Cole \& Lacey 1996; Lemson \& Kauffmann
1999).

We assume that the disk that forms in a halo has
mass $M_d$ related to the halo mass by
\begin{equation}\label{M_d}
M_d=m_d M_h\,,
\end{equation}
and has angular momentum $J_d$ related to
the halo spin by
\begin{equation}\label{J_d}
J_d=j_d J\,,
\end{equation}
where $m_d$ and $j_d$ give the fractions of mass and angular-momentum
in the disk. Assuming that the disk
has an exponential surface density profile and that
the dark halo responds to the growth of the disk adiabatically,
the disk scale-length $R_d$ can be written as
\begin{equation}\label{R_d}
R_d=\frac{1}{\sqrt 2}\left(\frac{j_d}{m_d}\right)\lambda{r_{200}}f_r\,,
\end{equation}
where $f_r$ is a factor that depends both on halo profile and on
the action of the disk (see MMW for details). As shown in MMW, for
a given halo density profile, $f_r$ depends both on $m_d$ and on
$\lambda_d\equiv (j_d/m_d)\lambda$, but the dependence on
$\lambda_d$ is not very strong. Thus, if $j_d/m_d$ is constant,
the log-normal distribution of $\lambda$ will lead to a size
distribution which is roughly log-normal.

\subsubsection{The bulge component}

 Our empirical knowledge about the formation of galaxy bulges
is still very limited (e.g. Wyse, Gilmore \& Franx 1997).
Currently there are two competing scenarios in the literature, one
is the merging scenario and the other is based on disk
instability.

 In the merging scenario, galaxy bulges, like elliptical galaxies,
are assumed to form through the mergers of two or more galaxies
(Toomre \& Toomre  1972). Subsequent accretion of cold gas may
form a disk around the existing bulge, producing a bulge/disk
system like a spiral galaxy (e.g. Kauffmann, White \& Guiderdoni
1993; Kauffmann 1996; Baugh, Cole \& Frenk 1996; Jablonka, Martin
\& Arimoto 1996; Gnedin, Norman \& Ostriker, 2000). In this scenario,
the formation of the bulge is through a violent process prior to the formation
of the disk, and so the properties of the bulge component are not
expected to be closely correlated  with those of the disk that
forms subsequently.

 In the disk-instability scenario, low-angular momentum
material near the centre of a disk is assumed to form a bar due to
a global instability; the bar is then transformed into a bulge
through a buckling instability (e.g. 
Kormendy 1989; Norman, Sellwood \& Hasan  1996; Mao \& Mo 1998;
van den Bosch 1998; Noguchi 2000). The first of these instabilities
is well documented through direct simulation, the second less so.
According to both $N$-body
simulations (Efstathiou, Lake $\&$ Negroponte 1982) and analytic
models (e.g. Christodoulou, Shlosman \& Tohline 1995) disks may
become globally unstable when
\begin{equation}\label{epsilon_m}
\epsilon \equiv\frac{V_m}{{(GM_d/R_d)}^{1/2}} < \epsilon_0\,,
\end{equation}
where $V_m$ is the maximum rotation velocity of the disk, and
$\epsilon_0\sim 1$. As discussed in MMW, for a disk in a NFW halo,
this criterion can approximately be written as $m_d > \lambda_d$.
Thus, for given $\lambda_d$, there is a critical value $m_{d,c}$
for $m_d$ above which the disk is unstable. If the overall stellar
mass fraction  $m_g$ (defined as ratio of total stellar mass to
total halo mass) is smaller than $m_{d,c}$, the disk is
stable and there is no bulge formation in this scenario.
In this case, $m_b=0$ and $m_d=m_g$. Here, $m_b$ is the bulge
fraction, which links the mass of the bulge to the halo mass,
\begin{equation}
M_b=m_bM_h\,.
\end{equation}
If $m_g>m_{d,c}$, we assume that the bulge mass
is such that the disk has $\epsilon=\epsilon_0$,
i.e. the disk is marginally unstable. In this case, $m_d=m_{d,c}$
and $m_b= m_g-m_{d,c}$. Note that the gravity of the bulge
component must be taken into account when calculating $\epsilon$.
To do this, we include a bulge component in
the gravitational potential following MMW. For given $m_g$
and $\lambda_d$, we then solve for $m_b$ and $m_d$ iteratively.

To proceed further, we assume the angular momentum of the bulge to
be negligible. There are two ways in which the bulge may
end up with little angular momentum: the first is that it formed
from halo material which initially had low specific angular momentum;
the second is that the bulge material lost most of its
angular momentum to the halo and the disk
during formation. In the first case, we assume that the
specific angular momentum of the final disk is the same as
that of the dark matter, so that $j_d= m_d$. In the second case, the final
angular momentum of the disk depends on how much of the bulge's
initial angular momentum it absorbs. Numerical simulations by
Klypin, Zhao \& Somerville (2002) suggest that angular
momentum loss is primarily to the disk for bulges that form
through bar instability. In general, we assume a fraction of $f_J$
of the bulge angular momentum is transferred to the disk
component, and so $J_d = (m_d + f_J m_b)J$. Thus the effective
spin parameter for the disk is
\begin{equation}\label{lambdanew}
\lambda_d =\lambda (1+ f_J m_b/m_d)\,,
\end{equation}
and we use this spin to calculate the disk size.

To complete our description of the disk component, we also need to
model the size of the bulge. Since current models of bulge
formation are not yet able to make reliable predictions about the
size-mass relation, we have to make some assumptions based on
observation. Observed galaxy bulges have many
properties similar to those of elliptical galaxies.  We therefore
consider a model in which galaxy bulges follow the same
size-mass relation as early-type galaxies.
Specifically, we assume that bulges with masses higher than
$2\times10^{10}\Msun$ have de Vaucouleurs profiles and have a size-mass
relation given by equation (\ref{fitERMass}). For less massive
bulges, we adopt exponential profiles and two models for the
size-mass relation. In the first model, low-mass bulges
follow a size-mass relation which is parallel to that
of faint late-type galaxies but has a lower zero point (so
that it joins smoothly to the relation for giant ellipticals at
$M=2\times10^{10}\Msun$), i.e.
\begin{equation}\label{RMb1}
\Log(R_e/\kpc)=\left\{\begin{array}{ll}
  0.56\Log(M_b)-5.54 & \mbox{for $M_b>2\times10^{10}\Msun$} \\
   0.14\Log(M_b)-1.21 &\mbox{for $M_b<2\times10^{10}\Msun$}
\end{array}\right. \,,
\end{equation}
where $R_e$ is the effective radius of the bulge.  This model is
motivated by the fact that dwarf ellipticals obey a
size-luminosity relation roughly parallel to that of spiral galaxies
(Kormendy \& Bender 1996; Guzman  et al. 1997; see Figures 7 to 9).
In the second
model, we assume a size-mass relation which is an extrapolation of
that for the massive ellipticals, i.e.
\begin{equation}\label{RMb2}
\Log(R_e/\kpc)=0.56\Log(M_b)-5.54\,.
\end{equation}
In this case, faint bulges are small and compact, like compact
ellipticals (Kormendy 1985; Guzman et al. 1997). For simplicity,
we do not consider the scatter in the size-mass relation in either
case.

With the mass and size known for both the disk and bulge
components,  we can obtain the surface density profile of the
model galaxy by adding up the surface density profiles of the two
components:
\begin{equation}\label{I(r)}
I(r)=I_d(r)+I_b(r)\,,
\end{equation}
from which one can estimate the half-mass radius for each
model galaxy.

\subsubsection{The value of $m_g$}

If all the gas in a  halo can settle to halo centre to form
a galaxy, then $m_g\sim \Omega_{\rm B, 0}/\Omega_0$. For the
cosmological model adopted here this would imply $m_g\sim 0.13$,
much larger than most estimates of
the baryon fraction in galaxies. In reality, not all the gas
associated with a
halo may settle into the central galaxy, because feedback
from star formation provides a heat source which may expel some
of it. Based
on such considerations, we consider a feedback model in which the
mass fraction $m_g$ in a halo of mass $M_h$ is
\begin{equation}\label{m_g}
m_g={m_0\over 1+ (M_h/M_c)^{-\alpha}}\,,
\end{equation}
where $M_c$ is a characteristic mass, $\alpha$ is a positive
index, and $m_0$ is a constant representing the mass fraction in
systems with $M_h\gg M_c$ (e.g. White \& Frenk 1991). We set
$m_0=\Omega_{\rm B,0}/\Omega_0=0.13$, so that $m_g$ is suppressed
for small haloes. Galaxy wind models suggest that the circular
velocity corresponding to $M_c$ is about $150\kms$, i.e. $M_c\sim
10^{12}\Msun$, and the value of $\alpha$ is $2/3$.
If the intergalactic medium is preheated
to a high entropy, then $\alpha$ is about $1$ (Mo \& Mao 2002).

\subsubsection{Specific models}

To summarize, there are four key ingredients in the scenarios
described above. The first is the feedback process which gives the
mass fraction $m_g$. We use the parameterized form given in
equation (\ref{m_g}) to model this process, and the model
parameters are $m_0$, $\alpha$ and $M_c$. The second is the
bulge/disk ratio $B/D$. This ratio is assumed to be either
uniform on the interval $[0,1]$ or given by the instability
criterion. The third is the amount of
angular momentum transfer between bulge and disk
components, as characterized by the parameter $f_J$ in
equation (\ref{lambdanew}). The fourth is the size-mass relation of small
bulges characterized by equation (\ref{RMb1}) or (\ref{RMb2}).

To consider these different possibilities, we
have chosen seven models as illustrations. In the following we
summarize these models in some detail; their parameters
are listed in Table 2.

\begin{table}
\caption{Parameters for the different models itemized in
section 4.1.4.}
\begin{tabular}{ccccccccc} \hline
Model &  $m_0$ & $\alpha$ & $M_c$ & B/D & $f_J$ & $R_e(M_b)$\\
\cline{1-7}
 I  & 0.05 &  0  & $1\times10^{12}$ & random & 0. & eq.(\ref{RMb1}) \\
 II & 0.13 &  0.67  & $1\times10^{12}$ & random & 0. & eq.(\ref{RMb1}) \\
 III& 0.13 &  0.67  & $1\times10^{12}$ &  disk instability & 1. & eq.(\ref{RMb1})  \\
 IV & 0.13 &  0.67  & $1\times10^{12}$ &   disk instability & 0.& eq.(\ref{RMb1}) \\
 V  & 0.13 &  0.67  & $1\times10^{12}$ &   disk instability & 0.5 &eq.(\ref{RMb1})\\
 VI & 0.13 &  1.    & $1\times10^{12}$ &  disk instability &0.5  & eq.(\ref{RMb1})\\
 VII& 0.13 &  0.67  & $1\times10^{12}$ &  disk instability &0.5  & eq.(\ref{RMb2})\\
\hline
\end{tabular}
\end{table}

\begin{itemize}
\item {\bf Model I}:
Here the mass fraction is chosen to be a constant
$m_g=0.05$, in contrast to the feedback model where
$m_g$ changes with halo mass, and the $B/D$ ratio is assumed
to be random in the interval $[0,1]$ . Disks are generated
according to the model described in subsection 4.1.1,
and bulges are assigned sizes according to equation (\ref{RMb1}).
 No angular-momentum transfer from the bulge component
to the disk is assumed.
\item {\bf Model II}:
Here $m_g$ is
assumed to follow equation (\ref{m_g}), with $m_0=0.13$,
$\alpha=2/3$, and $M_c=10^{12}\Msun$. Other assumptions are the
same as Model I.
\item {\bf Model  III}:
In this model $m_g$ is
assumed to follow equation (\ref{m_g}), with $m_0=0.13$,
$\alpha=2/3$, and $M_c=10^{12}\Msun$. Bulges are generated based
on disk instability. All of the initial angular momentum of the
bulge is assumed to be transferred to the disk, i.e. $f_J=1$. The
size-mass relation of the bulge component follows equation
(\ref{RMb1}).
\item {\bf Model IV}:
This model is the same as
Model III, except that there is no angular momentum transfer, i.e.
$f_J=0$.
\item {\bf Model V}:
This model is also the same as Model
III, except that half of the initial angular momentum of the bulge
material is transferred into the disk (i.e. $f_J=0.5$).
\item {\bf Model VI}:
 This model is the same as Model V, except that
$\alpha$ is assumed to be 1 instead of $2/3$.
\item {\bf Model VII}:
 This model is the same as Model V, except that
the size-mass relation of bulge is equation (\ref{RMb2}) instead
 of Eq. (\ref{RMb1})
\end{itemize}

\subsubsection{Model predictions}

\begin{figure}
\epsfysize=14.0cm \centerline{\epsfbox{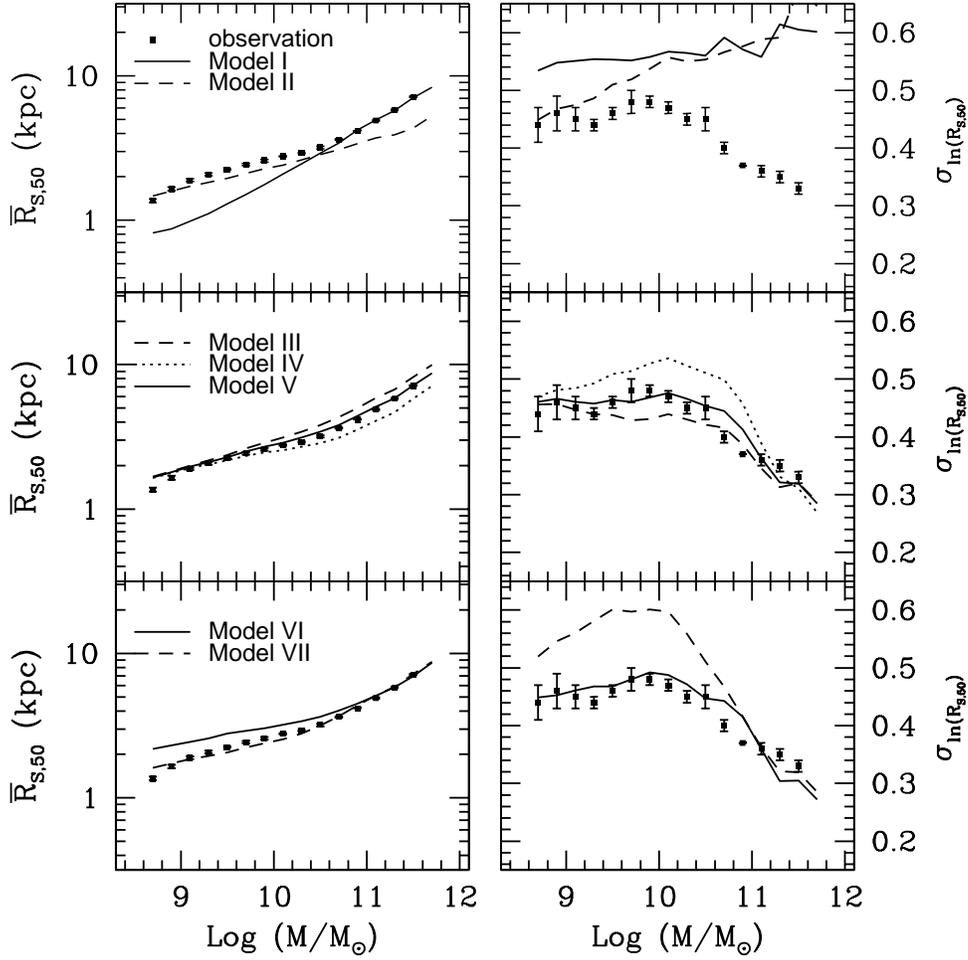}} \caption{The
median and dispersion of the distribution of half-mass radius of
spiral galaxies predicted by different models in comparison with
the observed distribution of the $z$-band \Sersic half-light
radius as function of stellar mass (Fig. 11). Observational
results are   shown only for galaxies with $n<2.5$. The models are
described in detail in section 4.1.4 and model parameters are
listed in Table 2.} \label{MCmodel}
\end{figure}

We use Monte-Carlo simulations to generate galaxy samples for each
of the models described above. To do this, we first use the
Press-Schechter (1974) formalism to generate 50,000 dark matter
haloes at redshift zero with masses (parameterized by circular
velocity $V_c$) in the range $35\kms<V_c<350\kms$ and with a
log-normal spin parameter distribution with $\bar{\lambda}=0.03$
and $\sigma_{\ln \lambda}=0.45$ (see equation \ref{p_of_lambda}).

We then use equations (\ref{r200})
-- (\ref{m_g}) to calculate the sizes and masses of the disk and bulge
for each galaxy. Finally, we combine the disk and bulge of each galaxy to
calculate its half-mass radius. As for the observational data, we sort
galaxies into stellar mass bins and calculate the median and
dispersion of the size distribution as functions of stellar mass.

Figure \ref{MCmodel} compares results for the seven models with the
observational data for the $z$-band \Sersic half-light radii of
late-type galaxies ($n<2.5$) as function of stellar masses.
  As one can see, if $m_g$ is assumed to be a constant,
(Model I), the predicted median size-mass relation is $\bar{R}\propto
M^{1/3}$, which is completely inconsistent with observations of low
mass galaxies. The predicted $\sigma_{\ln R}$ is also too large. When
$m_g$ is assumed to change with halo mass as suggested by the feedback
scenario (Model II), the predicted shape of $\bar{R}$ - $M$ is similar
to that observed for low mass galaxies, but the predicted median sizes
are too small for high mass galaxies. The predicted dispersion is also
too big for high mass galaxies. If bulges are assumed to form through
disk instability and if $f_J=1$ (Model III), the predicted scatter
follows the observations, but the predicted median sizes are too big
for massive galaxies. On the other hand, if no angular momentum is
transferred, i.e. $f_J=0$ (Model IV), the predicted median size is
smaller at the high mass end and the predicted scatter becomes
systematically higher. To match the observed behavior of the median
and the dispersion simultaneously, $f_J\sim0.5$ seems to be required
(Model V). Changing the value of $\alpha$ from $2/3$ to $1$ (Model VI)
gives a median size which is slightly too high for low-mass
galaxies. If the size-mass relation for small bulges is assumed to be
an extension of that for big ellipticals (Model VII), higher scatter
is predicted for low mass galaxies.

\begin{figure}
\epsfysize=10.0cm \centerline{\epsfbox{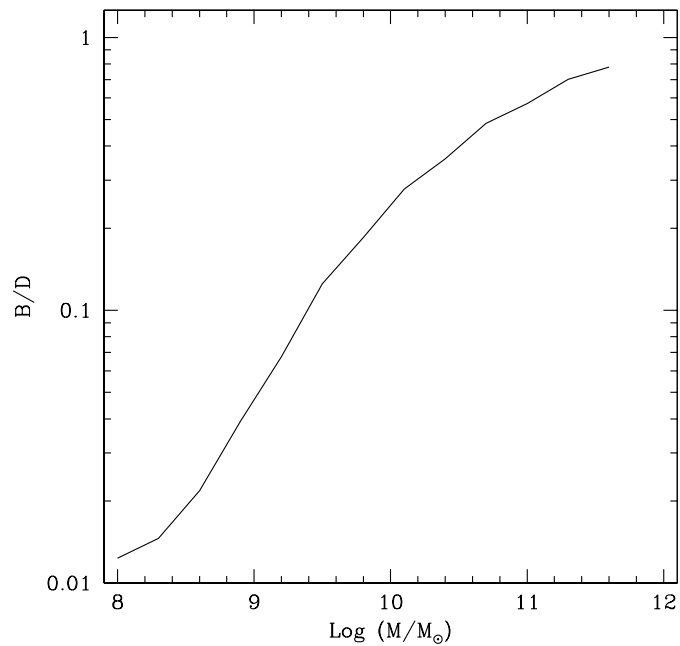}}
\caption{The prediction of Model V for the
average bulge/disk ratios for galaxies
with different stellar masses.}
\label{aBDM}
\end{figure}

Given that Model V reproduces the observed $\bar{R}$ - $M$
and $\sigma_{\ln{R}}$ - $M$ relations, it is interesting to
look at other predictions of the model. In this case, the
bulge/disk ratio depends on the mass of the halo. A
larger halo mass gives a larger value of $m_g$ and, in the
disk-instability model, this implies a larger bulge fraction.
Hence we expect $B/D$ to increase with galaxy mass. Figure
\ref{aBDM} shows the mean $B/D$ ratio as a function of $M$
according to Model V. The predicted trend is consistent with the
observed correlation between $B/D$ and galaxy mass (e.g. Roberts
\& Haynes 1994).

\begin{figure}
\epsfysize=10.0cm \centerline{\epsfbox{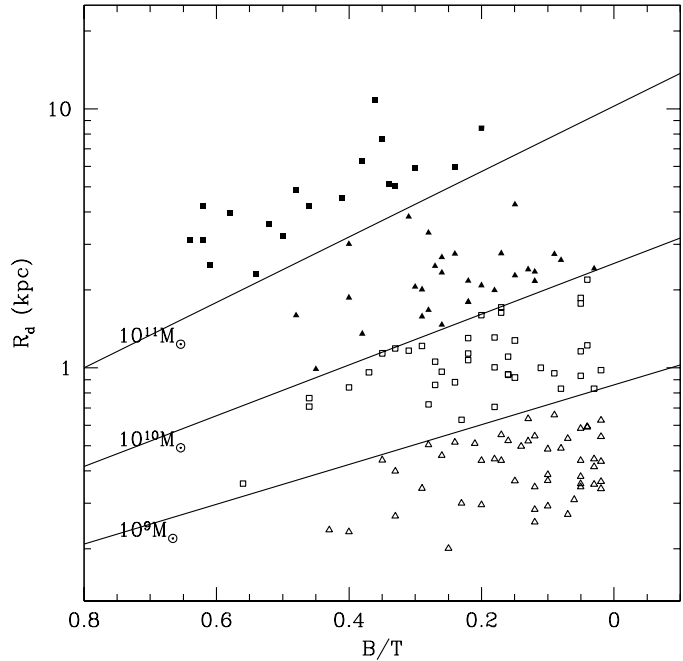}}
\caption{The prediction of Model V for the relation between
disk scale-length and bulge/total
mass ratio for late-type galaxies.
The points with different symbols represent galaxies in
different mass ranges (as shown by the solid lines).}
\label{RBD}
\end{figure}

The formation of the bulge depends on the properties of the disk in
the disk-instability-driven scenario. The bulge fraction ($B/T\equiv
m_b/m_g$) should therefore be correlated with disk size.  In Figure
\ref{RBD}, we show disk size as a function of $B/T$ for a number
of randomly selected Model V galaxies. The points of different type
represent galaxies of differing stellar masses; open triangles, open
squares, solid triangles and solid squares represent galaxies with masses
from $10^8\Msun$ to $10^{11}\Msun$ respectively. The rough separation
of galaxies according to mass is delineated by the solid lines in the
figure.  On average, galaxies with larger $B/T$ ratios have larger
disks. This is mainly due to the positive correlation between
$B/T$ and galaxy mass. For a given stellar mass, galaxies with large
$B/T$ ratios have smaller disks. This is consistent with the observational
results of de Jong (1996). If the bulge/disk ratio is assumed to be
random, as in Models I and II, disk size is independent of $B/T$.

\subsection {Early-type galaxies}

\begin{figure}
\epsfysize=14.0cm \centerline{\epsfbox{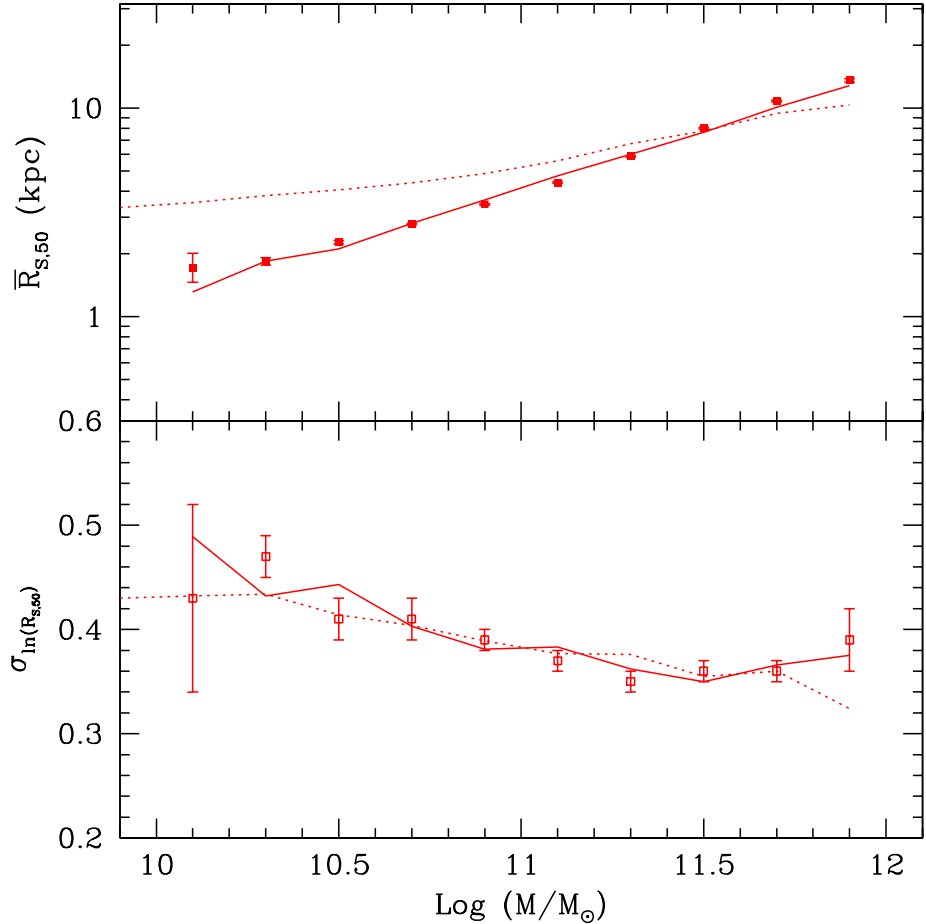}} \caption{
Model predictions for the size distribution of early type galaxies.
The solid lines assume ellipticals to be built by
repeated merging from a population of small
progenitors, while the dotted lines show a model
where each elliptical forms from the merger of two similar, late type
galaxies.
The observed size distribution of early type galaxies ($n>2.5$)
is reproduced from Fig. \ref{SRdisMassz} for comparison.}
\label{Emodel}
\end{figure}

Currently the most appealing model for the formation of elliptical
galaxies assumes they result from the merging of smaller systems.
Numerical simulations have shown that mergers of disk galaxies of
similar mass do indeed produce remnants resembling elliptical
galaxies (e.g. Negroponte \& White 1983; Hernquist 1992). However,
it seems unlikely that every elliptical is the remnant of a merger
between two similar spirals drawn from the observed local
population.  On the one hand, the stellar population of early-type 
galaxies is found to be so old that the typical 
star formation epoch must be at $z>2$ (e.g. Bernardi et al.
1998, 2003d; Thomas, Maraston \& Bender 2002). On the other 
hand, detailed modelling
of the merger histories of galaxies in a CDM universe suggests
that each elliptical obtains its stars from progenitors covering a
wide range in stellar mass, that the effective number of
progenitors increases weakly with the mass of the elliptical, that
the last major merging event is typically around redshift unity
but with a wide dispersion, and that the progenitors may have been
gas-rich, producing a substantial fraction of the observed stars
during merger events (Kauffmann 1996; Baugh et al. 1996; Kauffmann
\& Charlot 1998; Kauffmann \& Haehnelt 2000). Rather than treating
the detailed merger statistics of CDM models, we here contrast two
simpler models, one where ellipticals are built up by random
mergers within a pool of initially small progenitors, the other
where they form through a single merger of a pair of similar
``spirals''. As we will see, these pictures predict rather
different size-mass relations for the resulting population.
Consider two galaxies with stellar masses $M_1$ and $M_2$, and
corresponding half-mass radii $R_1$ and $R_2$, which merge to form a
new galaxy with stellar mass $M$ and size $R$. If we assume that all
of the stars end up in the remnant, then $M=M_1+M_2$. This is an
approximation, because numerical simulations show that a small amount
of mass typically becomes unbound as a result of violent potential
fluctuations during merging.  On dimensional grounds we can write the
total binding energy of the stars in each progenitor as $E_i= -C_i G
M_i^2 /R_i$ ($i=1, 2$), where $C_i$ depends on the density structure
of the galaxy in consideration.  In the absence of dark matter,
$C_i\approx 0.25$ for an exponential disk, while for a Hernquist
(1990) profile (for which the projected profile approximates the
$R^{1/4}$ law), $C_i\approx 0.2$. If we assume that the two
progenitors and the merger remnant all have similar structure, we can
write
\begin{equation}\label{merge}
\frac{M^2}{R}=\frac{M_1^2}{R_1}
+\frac{M_2^2}{R_2}+f_{\rm orb}\frac{M_1M_2}{R_1+R_2},
\end{equation}
where $f_{\rm orb}$ is a parameter which encodes the amount of energy
transferred from the stellar components of the two galaxies to the
surrounding dark matter as they spiral together.  The form of this
term is a simple model suggested by the expected asymptotic scalings
(see Cole et al. 2000) but unfortunately the appropriate value of
$f_{\rm orb}$ depends both on the structure of the galaxies and their
haloes and on the details of the merging process.  If we assume the
galaxies to have no dark haloes and to merge from a parabolic orbit,
then $f_{\rm orb}=0$. If the two progenitors are identical then in
this case $M=2M_1$ and $R=2 R_1$. It is easy to see that $R\propto M$
also for repeated mergers with these assumptions. Since our SDSS
results imply $R\propto M^{0.56}$, this simple model can be ruled out.

In order for two galaxies with the same mass ($M_1=M_2$) and radius
($R_1=R_2$) to merge to form a new galaxy with mass $M=2M_1$ and
radius $R=2^{0.56} R_1$, equation (\ref{merge}) requires $f_{\rm
orb}\approx1.5$. If we assume $f_{\rm orb}$ remains constant at this
value and repeat the binary merging $p $ times, each using remnant
from the previous time, then the mass and size of the remnants will
grow as $M=2^p M_1$ and $R= 2^{0.56p} R_1$.  Then $R\propto M^{0.56}$,
reproducing the observed relation.  Motivated by this, we consider a
model in which a giant elliptical is produced by a series of mergers
of small galaxies. We note that the observed masses ($\sim
10^{10}\Msun$) and half-mass radii ($\sim 1 h^{-1}\kpc$) of faint
early-type galaxies (see Fig. \ref{SRdisMassz}) are similar to the
masses and half-mass radii of Lyman-break galaxies observed at $z\sim
3$ (Giavalisco, Steidel \& Macchetto 1996; Lowenthal et al. 1997;
Pettini et al. 2001; Shu, Mao \& Mo 2001). These may perhaps be
suitable progenitors. We assume the progenitor population to have
masses $\sim 10^{10}\Msun$ and sizes given by a log-normal
distribution with $\bar {R}=1.3\kpc$ and $\sigma_{\ln{R}}=0.5$ (as
observed for faint ellipticals). We first use a Monte-Carlo method to
generate 100,000 progenitors. We randomly select two galaxies from the
progenitor pool and merge them to form a new galaxy according to
equation (\ref{merge}). After returning the new galaxy to the pool and
deleting its progenitors, we repeat this procedure many times.  Based
on the discussion above, we assume $f_{\rm orb}$ to be normally
distributed with mean $\bar{f}=1.5$ and a dispersion $\sigma_f$ to be
specified.  After 90,000 mergers, we obtain 10,000 galaxies with a
broad distribution of mass and radius.  For this sample we use the
maximum likelihood method described in Section 3 to estimate the
median and dispersion of the size distribution as functions of stellar
mass. The value of $\sigma_f$ is tuned to 1.35 so that the
predicted $\sigma_{\ln R}$ matches the observations for the most
massive galaxies. The model predictions so obtained are shown in
Fig. \ref{Emodel} as the solid curves, together with the observational
results. This simple model nicely reproduces the
observed size distribution for early-type galaxies.

For comparison, we have considered another model in which early-type
galaxies are produced by a single major merger between two present-day
spirals. We define a major merger to be one where the mass ratio of
the progenitors is larger than $1/3$. We use the observed late-type
galaxy population as progenitors, and randomly merge two of them to
form an early-type galaxy with size given by equation (\ref{merge}).
Here again $f_{\rm orb}$ is assumed to have a normal distribution with
mean 1.5 and dispersion 1.35. The results for $\bar{R}$ - $M$ and
$\sigma_{\ln R}$ - $M$ are shown as the dotted lines in Fig.
\ref{Emodel}. In this case, $\bar{R}$ is predicted to scale with $M$
in the same way as for the progenitors. This conflicts with
the observational results for early-type galaxies.  A possible
resolution may be that typical merger epochs are later for more
massive systems. Since disk galaxies are predicted to be larger and
lower density at later times (e.g. MMW; Mao, Mo \& White 1998) this results
in a steepening of the predicted dependence of $R$ on $M$. From the
scaling laws of MMW it easy to show that the typical cosmic time
at which merging occurs has to increase with elliptical mass
roughly as $t\propto M^{0.4}$ to reproduce the observed scaling
of size with mass. This slow dependence is perhaps compatible with the
expected dependence of formation time on halo mass in
hierarchical cosmologies (Lacey \& Cole 1993).

\section{Discussion and summary}

In this paper, we use a sample of $140,000$ galaxies from the SDSS to
study the size distribution of galaxies and its dependence on the
luminosity, stellar mass and morphological type of galaxies. This
database provides statistics of unprecedented accuracy.  These confirm
a number of previously known trends, for example, the approximately
constant surface brightness of luminous late-type galaxies (the
``Freeman disk''), the Kormendy relations between surface brightness
and luminosity for ellipticals, and a roughly log-normal form for the
size distribution function at fixed luminosity. We are able to
quantify these properties, and to show other relations that cannot be
seen in smaller samples. We find that, for late-type
galaxies, there is a characteristic luminosity at $M_{r,0}\sim -20.5$
(assuming $h=0.7$) corresponding to a stellar mass $M_0\sim
10^{10.6}\Msun$. Galaxies more massive than $M_0$ have median size
$\bar{R}\propto M^{0.4}$ and have dispersion in the size distribution
$\sigma_{\ln R}\sim 0.3$.  For less massive galaxies, $\bar{R}\propto
M^{0.15}$ and $\sigma_{\ln R}\sim 0.5$. The $\bar{R}$ - $M$ relation
is significantly steeper for early-type galaxies, with $\bar{R}\propto
M^{0.55}$, but the $\sigma_{\ln R}$ - $M$ relation is similar to that
of late-type galaxies.  Fainter than $M_r\sim -20$ the properties of
red galaxies are not a simple extrapolation of the relations for
bright early-type systems. These faint galaxies have low
concentrations and their half-light radii are almost independent of
luminosity. Brighter than $M_r\sim -20$ the mean surface brightness of
early-type galaxies also declines, so that systems near $M_r = -20$
have the highest values. In contrast, the average surface {\it
mass} densities of early-type galaxies are independent of luminosity
above $M_r\sim-20$.

We use simple theoretical models to understand the implications
of our observational results for galaxy formation.
We find that the observed $\bar{R}$ - $M$ relation for
late-type galaxies can be explained if the material
in a galaxy has specific angular momentum similar to that
of its halo, and if the fraction of baryons that form stars
is similar to that in standard feedback
models based on galactic winds. A successful model for the
observed $\sigma_{\ln R}$ - $M$ relation also requires the
bulge/disk mass ratio to be larger in haloes of lower angular
momentum and the bulge material to transfer part of its angular
momentum to the disk component. We show that this can be
achieved if the amount of material that forms a galactic bulge is
such that the disk component is marginally stable.

For early-type galaxies, the observed $\sigma_{\ln R}$ - $M$ relation
is inconsistent with the assumption that they are the remnants of
major mergers of present-day disks.  It may be consistent with a model
where the major mergers which formed lower mass ellipticals occurred at
earlier times and so involved more compact disks. The observed
relation is consistent with a model where early-type galaxies
are the remnants of repeated mergers, provided that the progenitors
have properties similar to those of faint ellipticals and that the
orbital binding energy is significant when two galaxies merge.

A number of issues remain unresolved in the present study.
First, the photometric errors of the half-light radii are
not considered in our analysis, which would finally convolved in
the derived dispersions of  the size distribution. Unfortunately,
an accurate assessment of error is difficult to make.
To test how significant this effect can be, we have done
Monte-carlo simulations including artificial errors in the 
measurements of the sizes. We found that if the measurement errors
are less than 10 percent (assuming a Gaussian 
distribution), the effect on the derived width of 
size distribution is negligible. Another uncertainty
is connected to the fact that our galaxy sample 
covers a non-negligible range in redshift while
galaxy sizes are based on half light radii in a fixed 
band in the {\it observational} frame rather than in 
a fixed band in the rest frame. However, this effect
should be quite weak because the sizes of galaxies are quite
independent of the wavelength as we have found (see also Bernardi
et al. 2003b). Yet another uncertainty may be caused 
by luminosity evolution of galaxies, which may affect  
the derived $R-L$ relation
(see e.g. Schade et al. 1996, 1997; Bernardi et al. 2003b). 
This effect should not be large in our results,
since most of our galaxies are located in a relatively
narrow redshift range ($0.05<z<0.15$). As a check, 
we have analyzed a sample which only includes galaxies 
with redshift $z<0.1$ and found negligible change 
in any of our results. Finally,
since faint ellipticals have light profiles similar to those of
disc galaxies, type classifications based on concentrations and
profile indices miss these objects.  Classifying according to
color suggests that these galaxies may have properties different
both from massive early-type galaxies and from spiral galaxies,
but it is unclear if all faint red galaxies in our samples are
ellipticals. To resolve this issue, we need a more accurate
indicator of morphological type. One way forward is to carry out
disk/bulge decompositions for a large number of galaxies. Such
work is underway in the SDSS Collaboration. With such
decompositions, we can study the properties of the disk and bulge
components separately, and so answer questions such as whether
bulges have similar properties to elliptical galaxies of the same
luminosity. 

\section*{Acknowledgments}

Funding for the creation and distribution of the SDSS Archive has been provided
by the Alfred P. Sloan Foundation, the Participating Institutions, the National
 Aeronautics and Space Administration, the National Science Foundation,
the U.S. Department of Energy, the Japanese Monbukagakusho, and the Max Planck
Society. The SDSS Web site is \url{http://www.sdss.org/}.

The SDSS is managed by the Astrophysical Research Consortium (ARC) for the
Participating Institutions. The Participating Institutions are The University of Chicago,
Fermilab, the Institute for Advanced Study, the Japan Participation Group,
The Johns Hopkins University, Los Alamos National Laboratory,
the Max-Planck-Institute for Astronomy (MPIA),
the Max-Planck-Institute for Astrophysics (MPA), New Mexico State University,
University of Pittsburgh, Princeton University, the United States Naval Observatory,
and the University of Washington.

{}
\end{document}